# Ride-sharing Determinants: Spatial and Spatio-temporal Bayesian Analysis for Chicago Service in 2022

**Mohamed Elkhouly[1*], Taqwa Alhadidi[2]**

[1]Assistant Professor, Department of Mathematics and Statistics, Cleveland State University, Cleveland, OH 44115, e-mail: m.elkhouly@csuohio.edu
[2]Associate Professor, Civil Engineering Department, Al-Ahliyya Amman University, Amman, Jordan 19111, e-mail: t.alhadidi@ammanu.edu.jo.

Corresponding author: Mohamed Elkhouly (e-mail: m.elkhouly@csuohio.edu).

**ABSTRACT** Ridesharing services have revolutionized transportation, transforming the manner in which individuals navigate urban environments. The estimation of ride-sharing demand is crucial for enhancing service utilization, reliability, and mitigating traffic congestion. In this research, we employ spatial and spatiotemporal Bayesian models to estimate ride-sharing demand in the 77 Chicago community areas. We investigate various explanatory variables, including demographic, socio-economic, transportation, and land-use features. The Bayesian models adjust for random and structured errors to account for spatial correlations. Additionally, we utilize spatially correlated priors for explanatory variables to improve model precision. Model diagnostics demonstrate the absence of residual spatial structure in the spatial regression, confirming effective management of spatial correlation. The spatiotemporal model yields a squared correlation of 0.95 between observed ride-shares and fitted values, indicating robust predictive power. Demographic factors, including population size and crime rate, emerge as the primary determinants positively influencing ride-sharing demand. Higher income levels, increased numbers of economically active citizens, and higher proportions of car-free households are associated with increased demand. Public transit availability and walkability also play significant roles in ride-sharing patterns in Chicago. Our study demonstrates that ride-sharing demand recovered in 2022 following COVID-19-related closures, particularly during weekends. This research provides planners and policymakers with insights to enhance first and last mile services, identify underserved areas, and enable ride-sharing to address public transportation gaps. Such knowledge has the potential to enhance mobility for individuals with disabilities and those residing in rural areas. This study, utilizing Bayesian models, offers deeper insights into ride-sharing demand, thereby contributing to improved urban transportation planning.

## I. INTRODUCTION

With rapid transportation development and innovative mobility solutions, ridesharing is now a widely used urban mobility mode [1]. It reduces congestion, serves last-mile trips, lowers emissions, and enhances network efficiency without additional infrastructure investments [2], [3], [4]. Key factors driving ride-sharing adoption include environmental awareness, ease of use, financial benefits, and social influence [5]. Challenges remain, such as reluctance to sacrifice private automobile flexibility and personal security concerns [6]. Integrating dynamic ride-sharing into local government portfolios has been explored to meet demographic needs like elderly services [7].

Sophisticated modeling techniques are essential for planning, optimizing, and implementing ride-sharing systems. Studies on analytics and AI-enabled personal information collection reveal perceived benefits and risks [8]. High service levels are needed to establish dynamic ride-sharing as a valid urban mobility alternative [9].

Research in Chicago examined user demographics, trip structures, and transportation impacts. 2020 data analysis identified user groups and their interactions with sociodemographic and built-environment factors [22]. Demographic factors influence preferences, with higher shared ride rates in areas with more young people in 2019 [23]. Fee structure fairness showed lower fees for shared trips compared to solo trips, highlighting economic aspects [24]. Services like UberPool reduce travel time between neighborhoods, improving transport efficiency [25]. Ride-hailing companies need to share detailed trip data with local agencies [26]. Insights into ride-sharing market dynamics were provided using real-world data [27].

Although various studies have modeled Chicago's ride-sharing, they have not addressed spatiotemporal characteristics [28]. Using spatial modeling, [29] employed a random-effects negative binomial regression to model daily origin-destination trips at the tract level, finding demand influenced by crime rates, employment density, parking availability, and fees. Similarly, [30] used a spatial error model (SEM) and geographically and temporally weighted regression (GTWR) to analyze one month's data, highlighting young people's different behaviors towards taxis on weekdays and weekends and the inverse relationship between transit and ride-hailing. GTWR's assumption of nonstationary coefficients and the use of one month's data raise reliability concerns, with significant coefficient variations suggesting model instability rather than robust nonstationarity of determinants.

This research advances understanding of urban mobility challenges in ride-sharing, using advanced Bayesian spatial and spatiotemporal modeling techniques. Our study analyzed determinants for ride-sharing across urban settings, providing insights applicable locally and globally. While most literature focuses on specific aspects of ride-sharing dynamics, this research offers a holistic view, integrating urban mobility factors, socioeconomic characteristics, land use, and temporal dynamics. These results contribute to urban transportation literature, emphasizing the complex interactions of urban form and transportation behavior, and offering validated models for city planners and policymakers to design efficient and sustainable urban transportation systems.

Bayesian spatial and spatiotemporal models were chosen for their ability to handle complex datasets with spatial and temporal dependencies. Classic regression methods often fail to capture nuanced patterns in urban mobility data, especially with irregular distribution or hierarchical structure. Bayesian models provide a flexible framework to integrate prior knowledge and quantify uncertainty, thus improving result reliability.

To summarize, the major contributions of this study are as follows:

1) Collect several determinants that affect ride-sharing demand in Chicago in 2022.
2) Model ride-sharing demand using spatial and spatio-temporal modeling techniques with one-year data. In this study, we developed two models, spatial and spatio-temporal, where we can use either model based on the planning purpose. If agencies are interested in checking the service using a spatial model, the spatial model is used. While the service is dynamic, spatiotemporal solutions can be used for short- and long-term solutions.
3) We scrutinized the spatiotemporal associations between ride-sharing trips and socio-demographic and transportation indices from both the spatial and spatio-temporal perspectives. Our findings provide valuable insights into the harmonized operation of traditional ride-sharing based on our understanding of travel patterns and usage determinants.
4) The Bayesian spatial CAR LeRoux model can incorporate spatially varying coefficients, meaning it allows the relationship between the dependent variable and predictors to vary across space. This flexibility is essential for capturing local variations in the effects of predictors on ride-sharing demand, providing a more nuanced understanding of spatial heterogeneity [1]. On the other hand, the non-Bayesian SEM, only accounts for spatial dependence in the error terms. In addition, SEM does not typically allow for spatial priors on the covariates and focuses primarily on modeling the error terms. Moreover, the SEM assumes that the relationship between the predictors and the dependent variable is constant across space, which may miss local variations in the effects of those predictors[2].
5) The Bayesian approach in estimation is flexible enough to account for various sources of uncertainty, such as realization measurement and stochastic error. Moreover, we consider hierarchical models in which the model parameters are learned from the data, that is, data talk. These models employ spatially structured and unstructured random effects and correlated time effects to handle the autocorrelations of the areal neighborhood structure and adjacent times. Therefore, the estimated parameters borrow strength and information both spatially and temporally simultaneously. Additionally, the considered models allow the predictors to vary spatially, as well as the response variable. Moreover, the Bayesian estimates come from the posterior predictive distributions averaging over the model's random effects. That makes the Bayesian approach is more realistic in addition to its capability to handle missing data with flexibility [1], [3].

The remainder of this paper is organized as follows. Section 2 reviews the literature on ride-sharing modeling. Section 3 describes the materials and provides details of the developed methods along with data collection. Section 4 presents the results, followed by a discussion in section 5. Finally, Section 6 concludes the study.

## II. Literature Review

Ride-sharing adoption is driven by user preferences, urban infrastructure, economic incentives, and technological advancements. Specifically, incentives significantly affect behavior, with demand patterns, algorithms, and urban settings shaping these incentives[4]. Additionally, demographic factors, such as low income and age, impact the willingness to use ride-sharing, highlighting socio-economic contexts[5]. Moreover, urban density and perceived value positively correlate with adoption, suggesting urban planners should consider spatial dynamics [6]. Despite these influences, ride-sharing's market share remains small compared to traditional ride-hailing, necessitating further research into adoption barriers [7]. Furthermore, psychological factors, including safety and social comfort, are significant, with psychological comfort being crucial[8] . In particular, comfort and safety are especially important for older adults using ride-hailing services [9]. Meanwhile, technological advancements in matching algorithms and routing optimization are vital for efficiency and appeal [10], [11]. Moreover, competitive pricing and lower costs for shared rides encourage participation and a shift from private vehicles to ride-sharing [5], [12]. Cultural attitudes and individual differences, such as travel distance and detour preferences, also shape behavior [13]. Finally, the COVID-19 pandemic has transformed ride-sharing, reducing private car dependence and carbon emissions [14]. While perceptions of shared modes will influence future transportation decisions in the post-pandemic era [15].

Unquestionably, ride-sharing has become more popular as a mode of transportation in urban areas, which has significant implications for policies and systems related to urban mobility. The examination of spatial variability in ride-sharing is currently a research topic [16] .This is particularly relevant when considering urban environments such as Chicago, where the dynamics of ride-sharing services may vary greatly between different regions. Furthermore, studies have investigated how ridesharing affects factors such as auto accidents and environmental benefits, providing important information about the broader effects of ridesharing models [17], [18]. New considerations arising from the development of ride-sharing models include multimodal ride-sharing and application of choice-based conjoint analysis to understand client preferences. These developments highlight the importance of considering different factors that affect the acceptability and effectiveness of ride-sharing models in urban environments. Furthermore, sentiment analysis from ride-sharing platform reviews has developed into a potent tool for learning about user preferences and experiences, thereby greatly advancing the subject of Kansei engineering [19].

Several factors affect ride-sharing demand, including social and economic factors and the ease of using services [20], [21]. Households with vehicles are more likely to use their cars than houses in high-density population areas; residences are more likely to use ride-sharing services [14], [22]. Social connections affect the demand for ride-sharing, as people are more likely to share rides with acquaintances [23].

Multiple studies have extensively examined many facets of ride-sharing services in Chicago, providing insights into user demographics, trip structure, and their influence on transportation networks. For example, research conducted by Rafi and Nithila (2022) examined data on ride-sharing trips in Chicago for 2020. The purpose of this study was to identify different user groups and understand how they interact with socio-demographic and built environment factors. This study offers significant insights into the user demographics of ride-sharing services in Chicago, which is essential for comprehending the industry and formulating efficient transportation policies [24]. Additionally, a study conducted by [25] emphasized the impact of demographic variables on ride-sharing preferences. This revealed that areas with a larger proportion of younger residents in Chicago in 2019 had a higher percentage of shared rides. The study findings emphasize the need to consider demographic variables when analyzing ride-sharing patterns in urban regions [25]. [26] investigated the fairness implications of ride-sharing pricing structures and found that Chicago applies distinct rates for individual and shared trips, with the latter subject to a reduced charge. This analysis of the cost structures of ride-sharing services in Chicago enhances the understanding of the economic factors associated with ride-sharing utilization in the city [26]. In addition, [27] showed that ride-hailing services, such as UberPool, can greatly decrease travel time between specific neighborhoods in Chicago. The study provides a practical perspective on how ride-sharing affects transportation efficiency in cities [27]. [28] conducted a study in Xiamen to examine the impact of ride-sharing on public transit. The findings revealed a substantial complementary effect of ride-sharing on subway systems, as supported by empirical evidence. While not specifically conducted in Chicago, the study provides useful insights into the potential synergies between ride-sharing services and public transit, which could be applicable to the transportation system in Chicago[28] . In addition, [29] emphasized the necessity of ride-hailing businesses sharing detailed trip data with local transportation agencies. They highlighted the importance of data openness and collaboration between ride-sharing companies and agencies in Chicago. [30] employed actual ride-sharing data from Chicago to examine ride-sharing patterns, offering empirical observations of the market dynamics of ride-sharing services in the city.

To simulate ride-sharing utilizing spatiotemporal and spatial analysis, it is crucial to consider multiple elements, including bike-sharing demand, ride-sharing trip data, ride-pooling, and the impact of ride-sharing on public transit. [16] suggested a method called spatial-temporal graph attentional LSTM to forecast bike-sharing demand. This method involves using many sources of data and a Graph Neural Network (GCN) to identify intricate spatiotemporal patterns. Adopting this method is essential for comprehending demand patterns and enhancing the efficiency of ride-sharing services in specific areas. K-Prototypes Segmentation Analysis was performed on a comprehensive dataset of ride-sharing trips, incorporating weather, transit, and taxi data to acquire a more profound understanding of the role of ride-sharing in the mobility system [31]. This analysis offers unique insights into the spatial and temporal patterns of ride-sharing, which are crucial for modelling the dynamics of ride-sharing. Moreover, the incorporation of ride-pooling into ride-sharing, as examined by[32], can enhance the intricacy of modelling and computation; however, it presents noteworthy advantages. Gaining insight into the influence of ride-sharing on its dynamics is essential for creating efficient models and maximizing the usage of shared rides [32]. A study conducted by [28] using regionally weighted regression revealed a positive relationship between the utilization of ride-sharing services and the number of individuals using public transit. This discovery emphasizes the interdependence of various modes of transportation and the necessity to consider their mutually beneficial impacts while constructing models for ride-sharing [28]. Furthermore, researchers have employed interpretable machine-learning models to analyze ride-sharing behavior based on data from the Chicago Transportation Network Company [30] . This observation is important for modelling ride-sharing behavior and optimizing techniques for ride-sharing journeys.

Ride-sharing has transformed urban mobility, reducing congestion, emissions, and the need for extensive infrastructure investments while enhancing the efficiency of transportation networks [33], [34]. Studies have consistently highlighted its role in serving last-mile trips and offering an alternative to private car usage [35]. The adoption of ride-sharing is influenced by environmental awareness, ease of use, financial benefits, and social factors [36].

Extensive research has been conducted to understand the factors influencing ride-sharing demand, particularly in urban settings like Chicago. [24] identified sociodemographic and built-environment factors that segment users, providing insights into targeted service enhancements. Similarly, [25] demonstrated how demographic profiles, especially younger populations, prefer ride-sharing, which aligns with broader trends of urban mobility preferences.

Economic considerations are crucial in shaping ride-sharing usage. [26] explored how differential pricing for solo and shared trips affects consumer choice, suggesting economic incentives can significantly modify user behavior. This is complemented by research from [27] , who found that ride-hailing services like UberPool effectively reduce travel times, making them an attractive option for cost-conscious users.

The spatiotemporal aspects of ride-sharing demand have been less frequently explored, presenting a gap our study aims to fill. Previous models have primarily focused on spatial correlations without integrating temporal dynamics. For instance, [37] used a random-effects model to analyze weekend ride-sharing demand, indicating time-based variations. [38] applied a geographically and temporally weighted regression, yet their study was limited by the short duration of data collection, highlighting the need for more robust longitudinal analyses.

The interaction between ride-sharing and public transit has been studied to understand their complementary roles.[28] highlighted a positive synergy between ride-sharing and subway usage, suggesting that ride-sharing can enhance public transit accessibility, particularly in underserved areas.

Our research builds on these foundational studies by employing a spatial and spatiotemporal Bayesian analysis to model ride-sharing demand across Chicago's community areas. By incorporating a broader range of demographic, socio-economic, and land-use characteristics, our models offer deeper insights into the dynamic interactions that influence ride-sharing. This approach not only addresses the identified research gaps but also enhances the practical applicability of our findings for urban planners and policymakers aiming to optimize ride-sharing integration into the urban mobility framework

Table I summarizes several studies on ride-sharing demand modeling. It compiles research findings from a selected group of articles published within the last five years, providing a systematic compilation of the main research contributions in this field. This coherent presentation offers readers a concise yet comprehensive understanding of the current state of knowledge on ride-sharing modeling.

TABLE I
SUMMARY OF RECENT STUDIES CONDUCTED FROM 2018 TO 2023

| Paper | Methods Used | Main Results | Limitations | Contributions |
|---|---|---|---|---|
| ([37] | Random-effects negative binomial (RENB) regression model. | The ridesharing demand is higher on weekends, in areas with higher incomes and population density. Moreover, the demand is higher on low-temperature and less-precipitation days, crime. | The study did not consider any specific spatial model. | Identifies determinants such as weather, socio-demographics, land-use, and crime. |
| [39] | In this article, the authors developed a statistical model of individuals' usage rates of ride-sharing services and found that lower income, older age, and the presence of small children in the household are characteristics that should be targeted as means of reducing transportation inequity. | - The paper explores factors influencing ride-sharing usage rates. - Socio-demographic and health-related variables were found to be significant. | N/A | N/A |
| [40] | In this article , the authors investigated factors influencing ride-sourcing usage in Bandung, Indonesia and found that two latent variables (comfort and reliability) are significant in ride-source choice. | - Ride-sourcing has a substantial impact on transportation in Bandung, Indonesia. - Factors influencing ride-sourcing usage include comfort, reliability, cost, and travel time. | N/A | - Waiting time is found to be insignificant in ride-sourcing choice. - All sociodemographic variables are insignificant except for house ownership. |
| [41] | In this paper , the authors investigated the main factors (individual sociodemographic, mobility-related characteristics, psychological attitudes, etc.) determining individuals' choices between ride-sharing and traditional taxis and found that people opened to technological innovation and with liberal thought tend to use ride-sharing services more often than taxis. | - Ride-sharing and traditional taxis provide similar on-demand transportation services. - Factors influencing users' choices include technology openness, liberal thought, and quality-of-service ratings. | N/A | N/A |
| [42] | In this paper , the authors investigated the influencing factors that impact ride-sharing service usage frequency and explore the potential similarities and differences among groups of population based on their primary usage purposes, including travelers' sociodemographic characteristics, reasons to choose ride-sharing services, and other behavioral characteristics. | - Investigates factors impacting ride-sharing service usage frequency in Shanghai - Identifies influencing factors and differences among groups based on primary usage purposes | N/A | N/A |
| [43] | In this article, the authors explored the travel behavior of ride-sourcing users and those users' socio-demographic characteristics as well as perception of the usefulness based on the users' previous modes of transport before ride-Sourcing existed. | - Study explores travel behavior and perception of ride-sourcing usefulness in Bandung City. - Substitution from public transport for younger travelers, private transport for infrequent and higher-income travelers. | - Research shows ride-sourcing substitutes for both private and public transport. - Study explores travel behavior and perception of usefulness based on previous modes. | - Substitution from public transport exists for younger travelers.<br>- Substitution from private transport is associated with infrequent and higher-income travelers. |
| [44] | In this article, the authors explored how ride-sharing adoption rate varies across space and what factors are associated with these variations and revealed nonlinear patterns can help transportation professionals identify neighborhoods with the greatest potential to promote rideshitting. | - Study explores ride-sharing adoption rate and factors associated with variations. - Socioeconomic, demographic, travel-cost, and built-environment factors influence ride-sharing adoption. | N/A | N/A |
| [45] | In this article, a robust data-driven understanding of how ride-sourcing demand is affected by the built environment, after controlling for socioeconomic factors, was provided by having unique access to Uber trip data in 24 diverse U.S. regions. | - Examines how built environment affects demand for Uber ride-sourcing services - Positive correlation with population, employment, activity density, land use mix, transit stop density; negative correlation with intersection density and destination accessibility | N/A | N/A |

## III. Methodology

Our methodological contribution lies in the integration and enhancement of spatial and spatiotemporal Bayesian statistical models. These models have been adapted to more accurately capture the complexities of urban ride-sharing dynamics, incorporating both structured and unstructured random effects to account for area-specific and temporal variations. This approach represents an improvement over traditional regression analyses by allowing for a more nuanced understanding of the factors influencing ride-sharing demand across different times and locations.

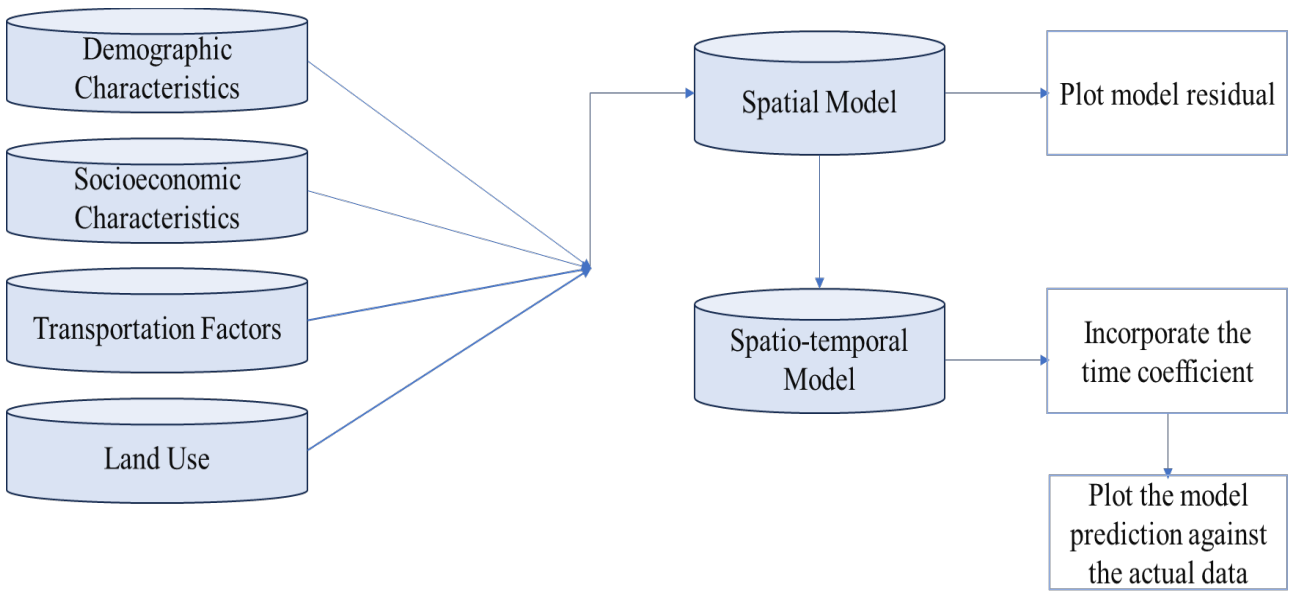

**FIGURE 1. Research Methodology**

Here, we model ride-sharing demand in Chicago using spatial dependencies, temporal dependencies, and exogenous dependencies, in ride-sharing demand forecasting. In this section, we first present the study framework, followed by a brief description of the data used, model building and checks, and the EDA. Figure 1 illustrates the research methodology.

In this study, we utilize two Bayesian frameworks to model ride-sharing in Chicago for the year 2022. Our investigation encompasses ride-sharing data across 77 community areas in Chicago, accessible at (https://data.cityofchicago.org/Transportation/Transportation-Network-Providers-Trips-2022/2tdj-ffvb/about_data).
The predictors we collected fell into three groups: socioeconomic, transportation, and land-use factors.
The first group of predictors includes various demographic and socioeconomic factors across the 77 Chicago community areas. These predictors are the 2020 population size, average household size, median age, median income, proportion of the population with a bachelor's degree or higher, percentage of the economically active population, percentage of residents with no vehicles, and registered crimes in 2022 in each area.

The second group comprises transportation factors that potentially influence ride-sharing demand in Chicago. This group includes the walkability index of residence and job locations (categorized as low, moderate, and high) and transit availability index (similarly categorized). The walkability index considers factors such as sidewalks, amenities within walking distance, population/employment density, bicycle/pedestrian crashes and fatalities, and physical characteristics. The transit availability index is based on factors such as the frequency of transit services, proximity to transit stops, activities reachable without a transfer, and pedestrian friendliness. We specifically consider moderate and high percentages of walkability and transit indices.

The third group focuses on general land use categories and their corresponding percentages. We examined the percentages of mixed, commercial, institutional, and industrial land use for each of the 77 areas. Additionally, for parking access, we use the open space per 1000 residents that considers the neighborhood and the community accessible by residents living within 0.5 and 1 mile respectively.

Meanwhile, the demographic predictors we consider, such as population size, median age, and others, are not expected to change significantly over a two-year period. Similarly, indices like walkability and transit availability are not likely to fluctuate dramatically within such a short timeframe. Therefore, for this research we assume that there are no dramatic changes in the spatial distribution of our explanatory variables that may affect our results significantly. However, this assumption may have been a possible source of error in the models.

The rideshares and registered crimes data are not time-invariant, in contrast to the other variables. Table II summarizes the different variables and convenient transformations that we employ when necessary. We consider the log transformation of rideshares, total population, median income, and crime. We consider the logit transformation for each of the economically active ratios and bachelor/graduate percent. This makes the rideshares data look Gaussian and achieves linearity in the relationship with the predictors.

We obtain the record of crimes from the Chicago data portal available at (https://data.cityofchicago.org/Public-Safety/Crimes-2022/9hwr-2zxp/about_data). All other predictors from Chicago community data snapshots are available at (https://www.cmap.illinois.gov/data/community-snapshots).

Our study utilized three main types of factors: demographic variables, economic indicators, and transportation infrastructure data. Each set of data was sourced from reputable databases to ensure accuracy and reliability.

1. **Demographic Variables:**
    - **Source:** U.S. Census Bureau
    - **Time Range:** Data collected from the 2020 Census.
    - **Preprocessing Steps:** Data were normalized to account for population differences across Chicago's community areas. Missing values, primarily from underreported sectors, were imputed using regression techniques based on nearby geographical areas with similar socioeconomic profiles.

2. **Economic Indicators:**
   - **Source:** Chicago Data Portal
   - **Time Range:** Economic data from 2019 to 2021, to capture pre- and post-pandemic economic conditions.
   - **Preprocessing Steps:** Economic data were adjusted for inflation to ensure consistency over the different years. Seasonal adjustments were also made to reflect typical economic cycles, using a standard seasonal decomposition technique.
3. **Transportation Infrastructure Data:**
   - **Source:** City of Chicago's Department of Transportation
   - **Time Range:** Continuous updates with the latest data pull in 2022.
   - **Preprocessing Steps:** Geographic data were standardized to a common coordinate reference system. Data cleaning involved removing outliers and correcting data entry errors identified through validation checks against known infrastructure maps.

**Data Integrity and Validation:** To ensure the integrity of our data, we conducted extensive validity checks, including comparisons with secondary sources and feedback from local authorities. This helped in refining our datasets to accurately reflect the current urban dynamics of Chicago.

### *A. Statistical models*
We employ two Bayesian frameworks to model rideshares across 77 Chicago areas, utilizing the three groups of predictors. The first model is a spatial regression of the total number of rideshares in 2022. We then consider the statistically significant factors from the first model to proceed with our analysis. The second model is spatiotemporal regression for daily rideshares throughout 2022. Below are descriptions of these models.

### *B. Spatial Regression of Total Rideshares in 2022*
To identify the statistically significant determinants for rideshares demand in Chicago, we fit a regression model with spatial dependence. Specifically, we employ the Conditional Auto-Regression (CAR) model proposed by [46]. In this context, the CAR model serves as a prior distribution for the spatial effects across 77 areas. We fit the adapted version of Leroux et al. (2000) using the Bcartime function in the bmstdr R package [3]. Equation (1) describes the hierarchical Leroux model, which accounts for the spatial autocorrelation of the areal data.

$$Y_k | \mu_k \sim Normal(\mu_k, \nu^2)$$

$$\mu_k = X_k^T \beta + \phi_k$$

$$\beta \sim Normal(\mu_\beta, \Sigma_\beta)$$

$$\nu^2 \sim Inverse - Gamma\ (a_1, b_1)$$

$$\phi_k | \Phi_{-k}, \boldsymbol{W}, \tau^2, \rho \sim Normal\ \left(\frac{\rho \sum_{i=1}^{K} \omega_{ki}\, \phi_i}{\rho \sum_{i=1}^{K} \omega_{ki} + 1 - \rho}, \frac{\tau^2}{\rho \sum_{i=1}^{K} \omega_{ki} + 1 - \rho}\right)$$

$$\tau^2 \sim Inverse - Gamma\ (a_2, b_2)$$

$$\rho \sim Uniform(0, 1)$$

Here, $\mu_k$ represents the expectation of $Y_k$ , where $x_k = (1, x_{k1}, \ldots, x_{kp})$ is the vector of covariates for the $k^{th}$ areal unit, and $k = 1, \ldots, K$ is the non-overlapping areal unites. The non-spatial error variance, $\nu^2$, and spatial variance, $\tau^2$, both have inverse gamma priors. For the p-regression coefficients $\beta = (\beta_1, \ldots, \beta_p)$, a multivariate Gaussian prior is employed with a mean $\mu_\beta$ and diagonal covariance matrix $\Sigma_\beta$. $\Phi = (\phi_1, \ldots, \phi_K)$ are the spatial random effects with CAR prior where $\rho$ governs the spatial smoothness. When $\rho = 0$, the model represents independent random effects while $\rho = 1$ implies that the model fully captures spatial dependence. The spatial correlation parameter $\rho$ has a uniform (0,1) prior. Lastly, W is a non-negative symmetric binary matrix representing the neighborhood structure based on geographical contiguity, where $\omega_{ki} = 1$ if the areal units $(S_k, S_i)$ share a common border.

### C. *Spatio-temporal Regression of the Daily Rideshares in 2022*
The second model analyzes daily rideshares by utilizing a spatio-temporal Bayesian model to account for both temporal and spatial correlations simultaneously. We adopt a framework with separable spatial and temporal dependence, featuring an autoregressive temporal dependence of order two (AR2). This model characterizes the spatiotemporal pattern in expected rideshares with a single set of spatially and temporally autocorrelated random effects proposed by Rushworth et al. (2014) [47] as in Equation (2).

$$Y_{kt} | \mu_{kt} \sim Normal(\mu_{kt}, \nu^2)\ for\ k = 1, \ldots, K\ and\ t = 1, \ldots, T$$

$$\mu_{kt} = X_{kt}^T B + \psi_{kt}$$

$$B \sim Normal(\mu_B, \Sigma_B)$$

Where $\mu_{kt}$ represents the expectation of $Y_{kt}$ and B is a vector of the p coefficients. The index $k = 1, \ldots, K$ spans non-overlapping areal units, and the index $t = 1, \ldots, T$ corresponds to the time point. The vector $x_{kt}$ of length p, captures the realization of the p predictors for areal unit k and time point t. Additionally, $\psi_{kt}$ represents spatio-temporally autocorrelated random effects. For more details on the model presented in Equation (2), refer to CARBayesST by Duncan Lee et al. (2018) [48]. The model was fitted using the BcarTime function in the bmstdr R package.

TABLE II
VARIABLE SUMMARY STATISTICS

| Variable | Transformation | Time Invariant | Min | Mean | Median | Max |
|---|---|---|---|---|---|---|
| Total Ridesharing 2022 | Log | No | 10.39 | 12.77 | 12.67 | 16.14 |
| Population 2020 | Log | Yes | 7.835 | 10.241 | 10.297 | 9.413 |
| Total crimes 2022 | Log | No | 5.71 | 7.73 | 7.692 | 9.413 |
| Household Avg Size | None | Yes | 1.556 | 2.527 | 2.501 | 3.816 |
| Median Income | Log | Yes | 9.698 | 10.395 | 10.969 | 11.864 |
| Bachelor and Graduate Percent | Logit | Yes | -2.628 | -0.750 | -0.928 | 1.778 |
| Economically Active | Logit | Yes | 0.773 | 2.278 | 2.325 | 4.014 |
| Median Age | None | Yes | 24.49 | 36.75 | 36.20 | 48.48 |
| No Veichle Proportion | None | Yes | 0.037 | 0.243 | 0.212 | 0.548 |
| Open Space /1000 Residents | None | Yes | 0.140 | 2.902 | 2.010 | 15.153 |
| Mixed Use Percent | None | Yes | 0 | 0.011 | 0.009 | 0.049 |
| Commercial Use Percent | None | Yes | 0.004 | 0.054 | 0.043 | 0.281 |
| Institutional Use Percent | None | Yes | 0.005 | 0.063 | 0.046 | 0.301 |
| Industrial Use Percent | None | Yes | 0 | 0.243 | 0.027 | 0.307 |
| Transit Low PCT | None | Yes | 0 | 0.005 | 0 | 0.256 |
| Transit Moderate PCT | None | Yes | 0 | 0.02 | 0 | 0.978 |
| Transit High PCT | None | Yes | 0.0127 | 0.974 | 1 | 1 |
| Walkability Low PCT | None | Yes | 0 | 0.0354 | 0.0 | 0.734 |
| Walkability Moderate PCT | None | Yes | 0 | 0.04037 | 0 | 0.778 |
| Walkability High PCT | None | Yes | 0 | 0.9242 | 1 | 1 |

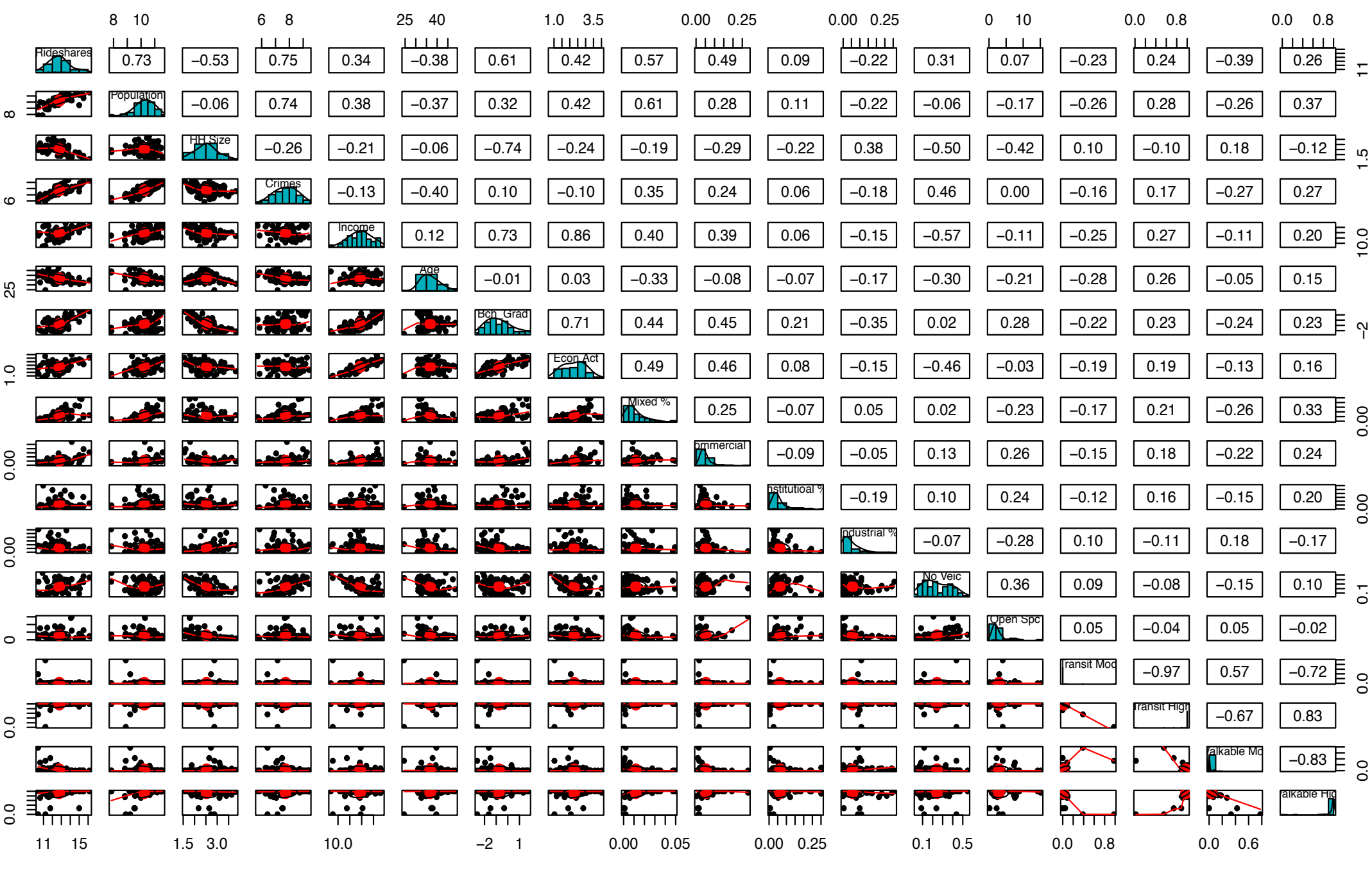

FIGURE 2. Matrix plot of the response and explanatory variables with Pearson's correlation

## IV. Results

### *A. Exploratory Data Analysis (EDA)*

After implementing the transformations outlined in Table (2), we calculate the correlation between the total rideshares and various explanatory variables. Figure (2) illustrates a strong positive correlation between rideshares and both population size and the number of crimes (r=0.71 and 0.75, respectively).

Additionally, there is a moderate positive correlation between rideshares and the percentage of residents with a bachelor's or graduate degree (r=0.61), the percentage of economically active (r=0.42), and the corresponding percentage of mixed and commercial land use (r=0.57 and 0.49). Moreover, there is a weak positive correlation between rideshares and the median income (r=0.34) and high transit index (r=0.24). On the other hand, some predictors exhibited a negative correlation with rideshares across the 77 Chicago community areas, including average household size (r=-0.53), median age (r=-0.38), and moderate walkability index (r=-0.39).

Despite the significant correlations observed, there is clear evidence of multicollinearity among predictors. For instance, there is a very strong correlation between the transit index levels (r=-0.97) and walkability index levels (r=-0.83). Similarly, there are high correlations between income and both the percentage of residents with a bachelor's or graduate degree (r=0.73) and the economically active percentage (r=0.86). Furthermore, there is a positive correlation between population size and percentage of mixed land use (r=0.61). Consequently, we do not anticipate all predictors correlated with rideshares to be statistically significant in our spatial regression model. Additionally, a variable selection procedure is necessary to identify the model with the most significant determinants of rideshares.

We proceed visualizing the spatial realizations of the different variables in Figure 3. Upon visual inspection, strong spatial dependence is apparent for each of the response and potential explanatory variables, except for the transit indices (moderate and high). This observation is further supported by the estimated Moran's index (Moran, 1950), which confirms significant spatial dependence. Thus, it is imperative to account for this spatial correlation when investigating rideshares determinants for 77 Chicago community areas. Furthermore, treating rideshare data as independent observations would lead to an overestimation of variance components, potentially swamping crucial predictors.

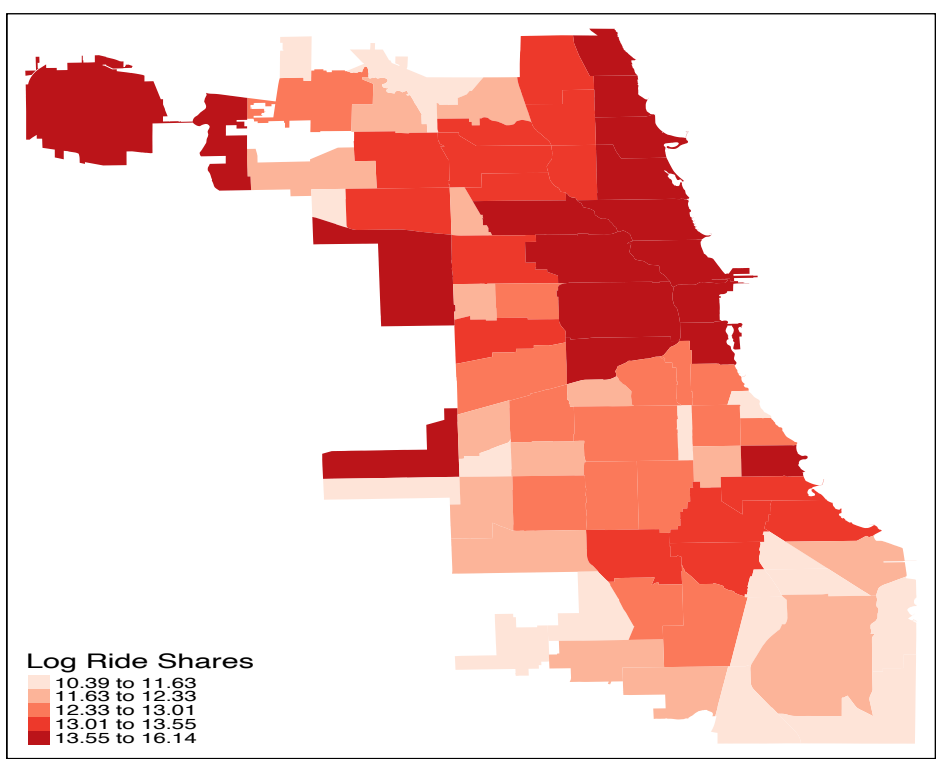

Moran's-Index = 0.460 (p=0.001)

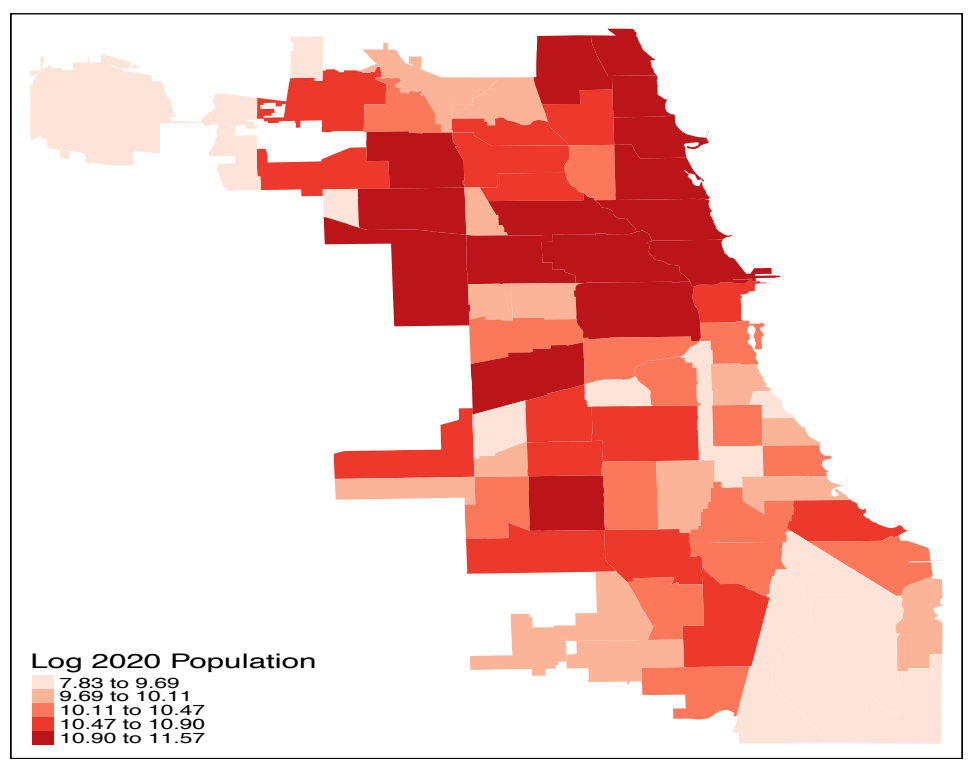

Moran's-Index =0.325 (p=0.001)

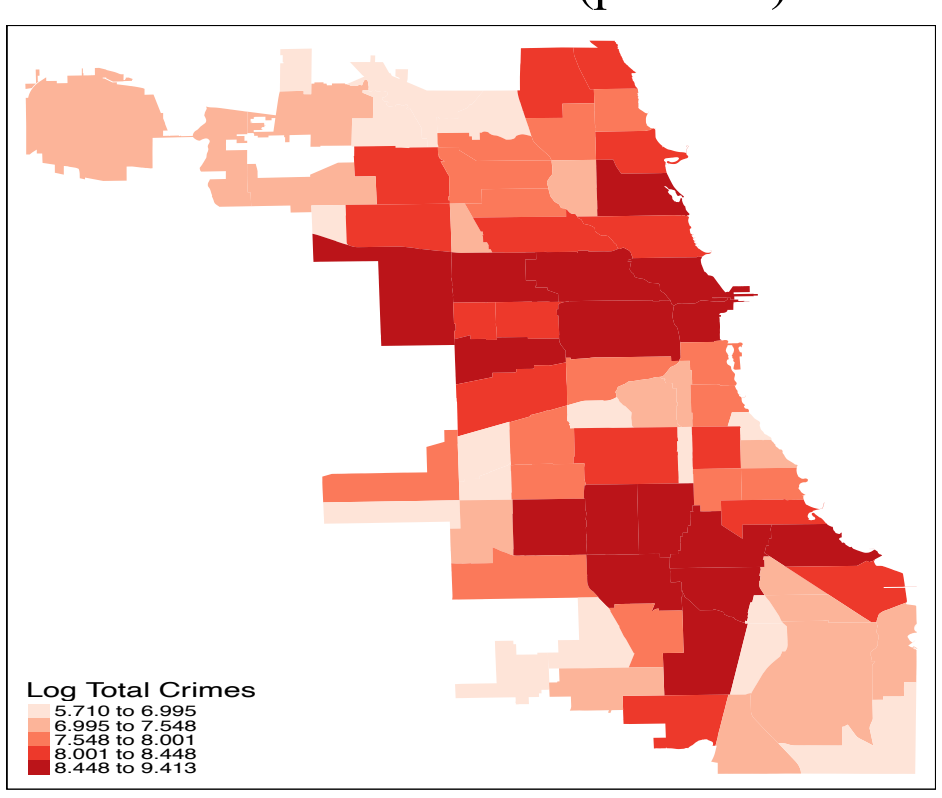

Moran's-Index = 0.269 (p=0.001)

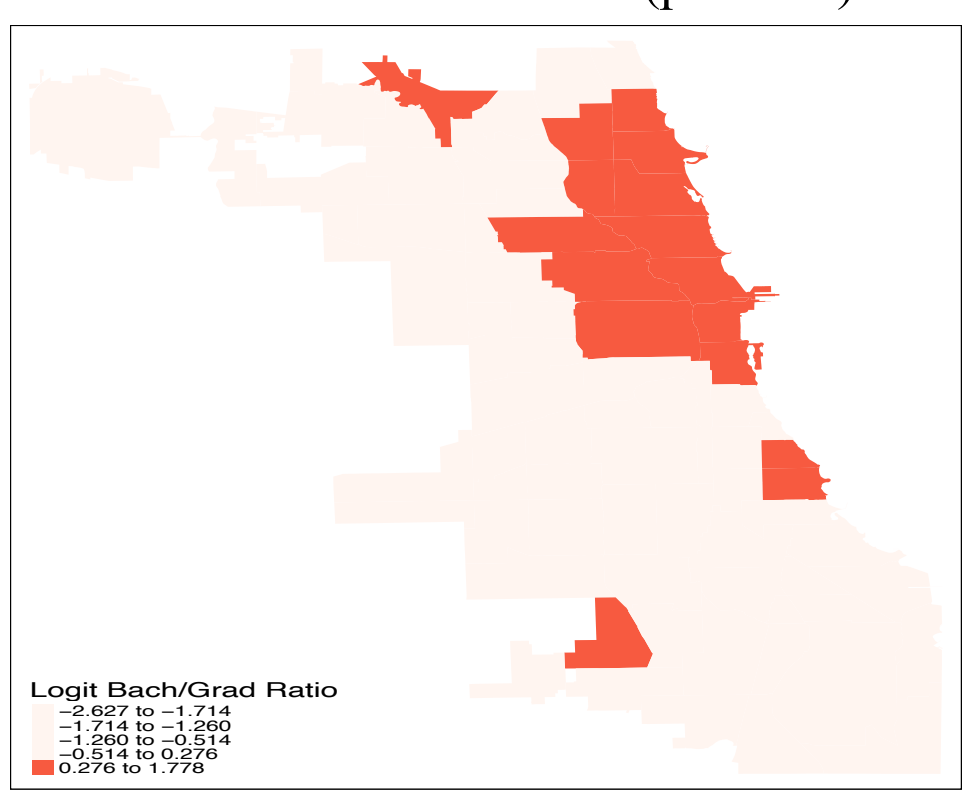

Moran's-Index = 0.661 (p=0.001)

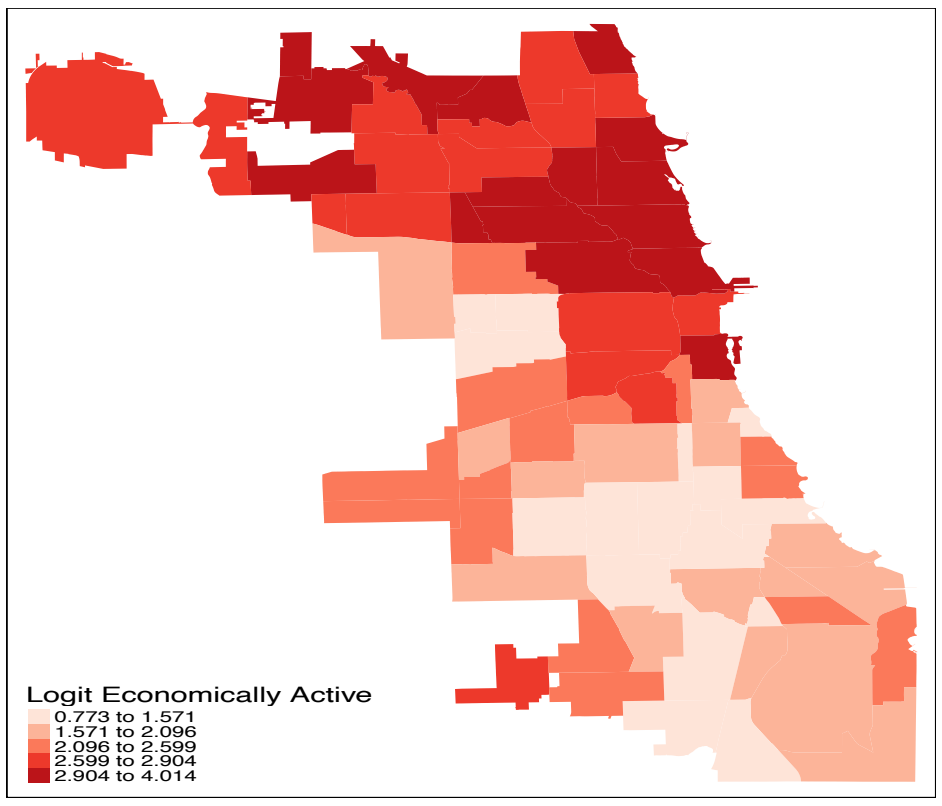

Moran's-Index = 0.621 (p=0.001)

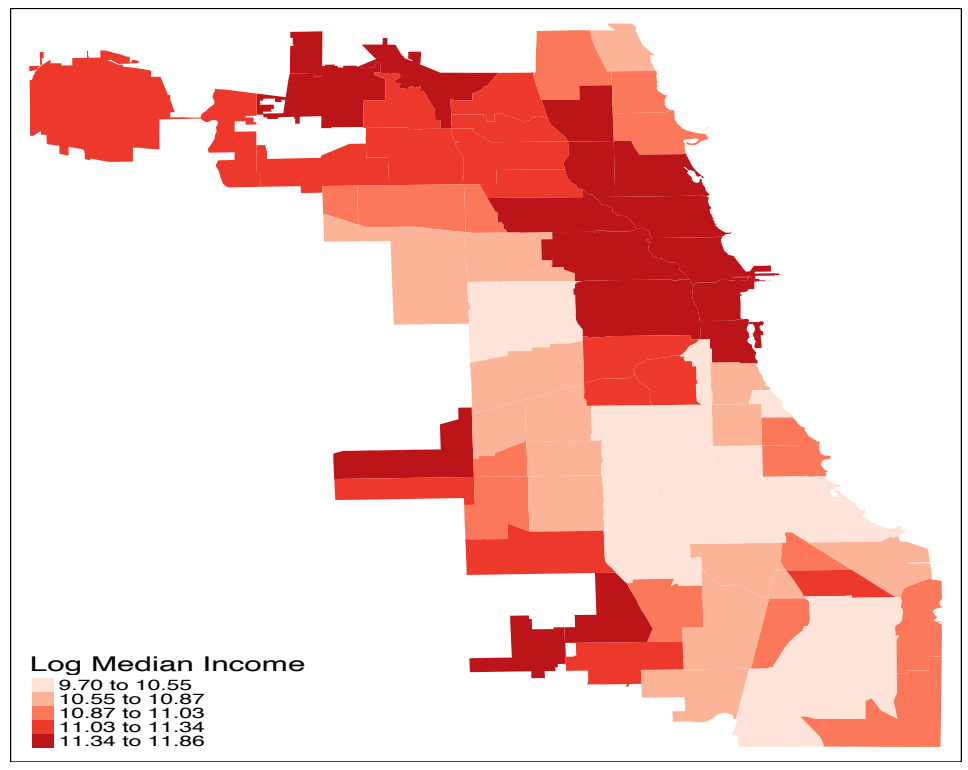

Moran's-Index = 0.552 (p=0.001)

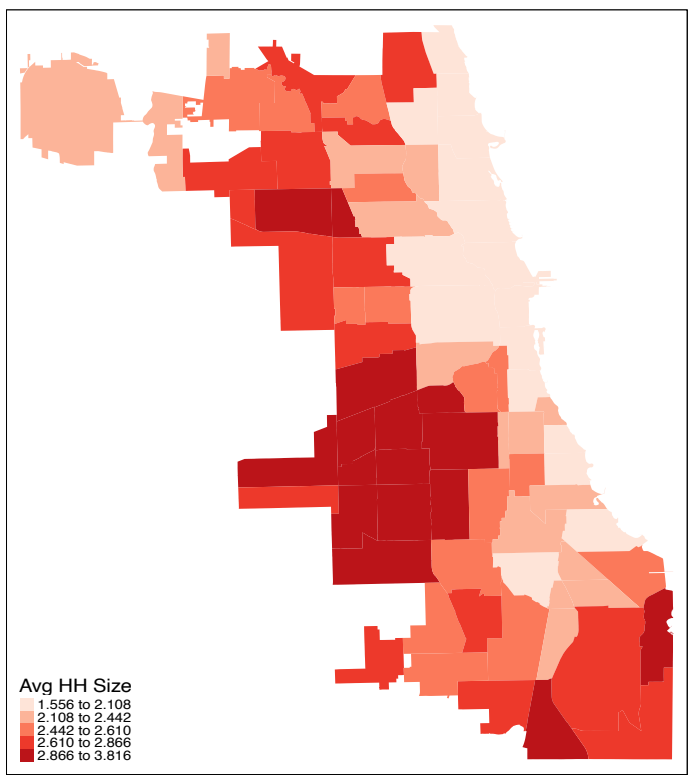

Moran's-Index =0.336 (p=0.001)

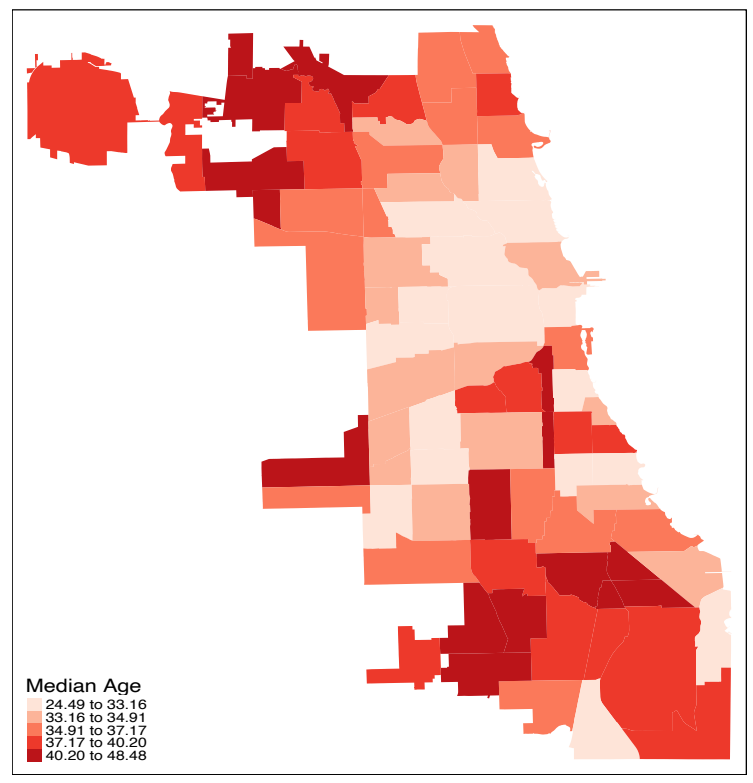

Moran's-Index = 0.728 (p=0.001)

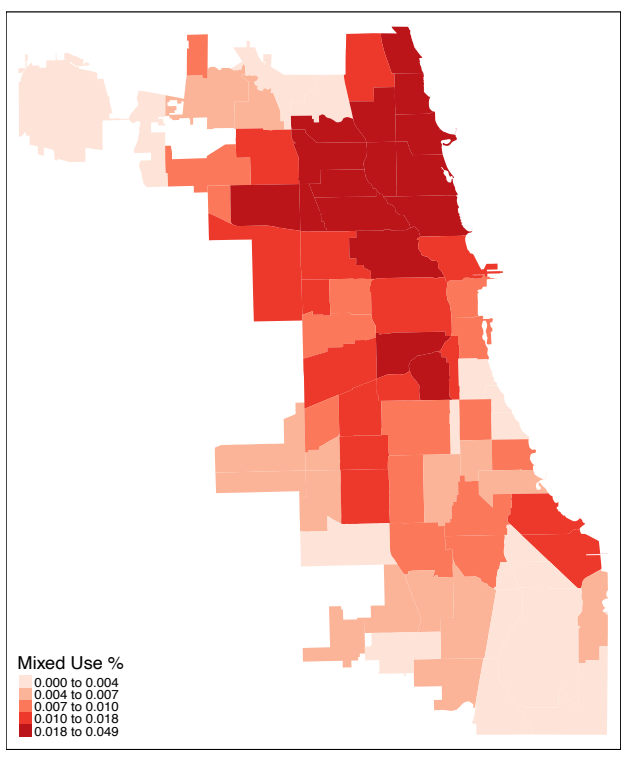

Moran's-Index = 0.610 (p=0.001)

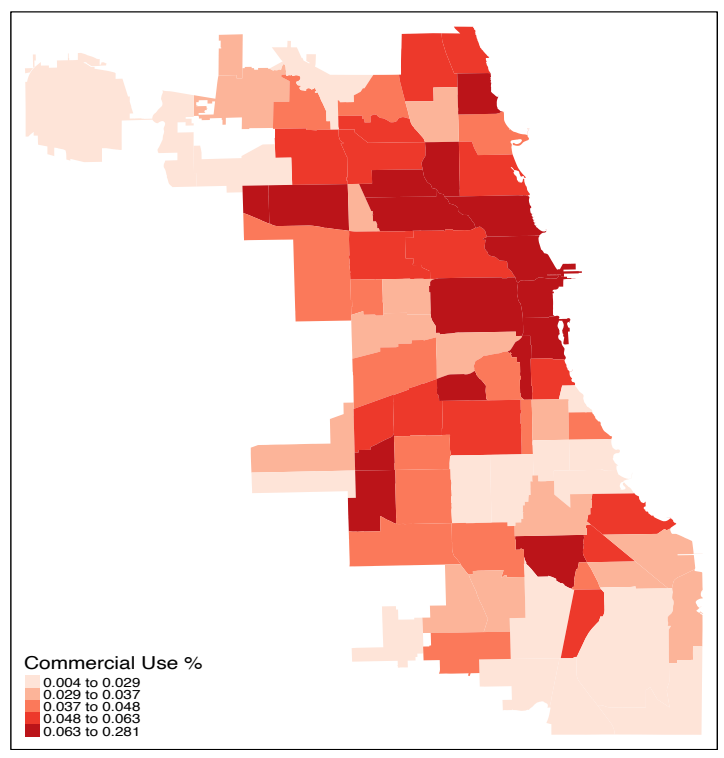

Moran's-Index = 0.367 (p=0.001)

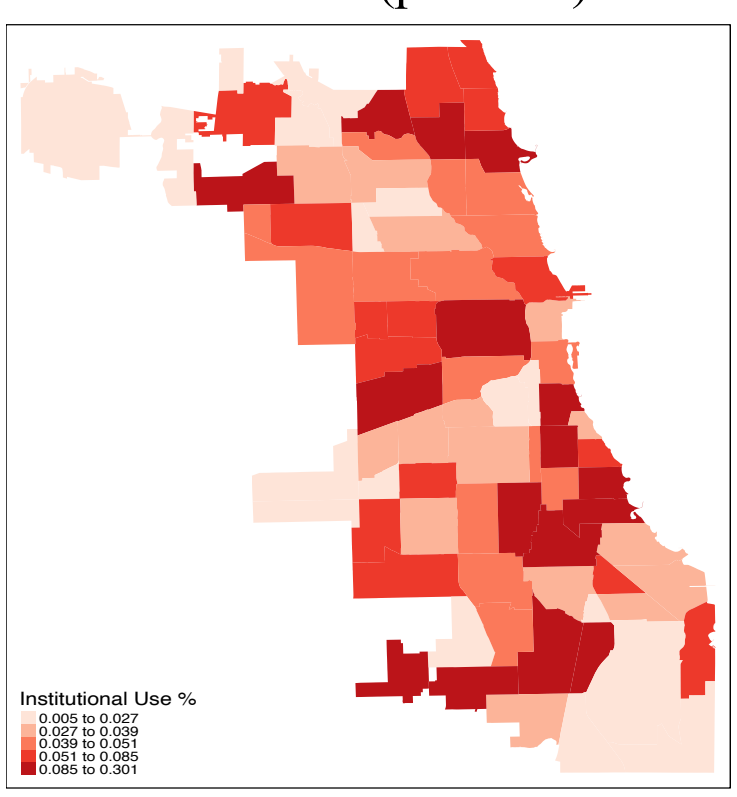

Moran's-Index = 0.783 (p=0.001)

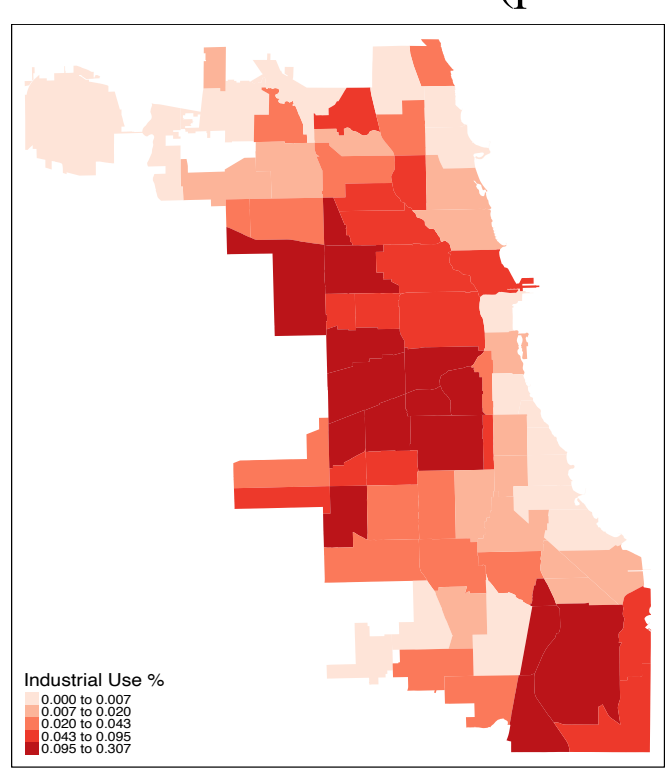

Moran's-Index = 0.378 (p=0.001)

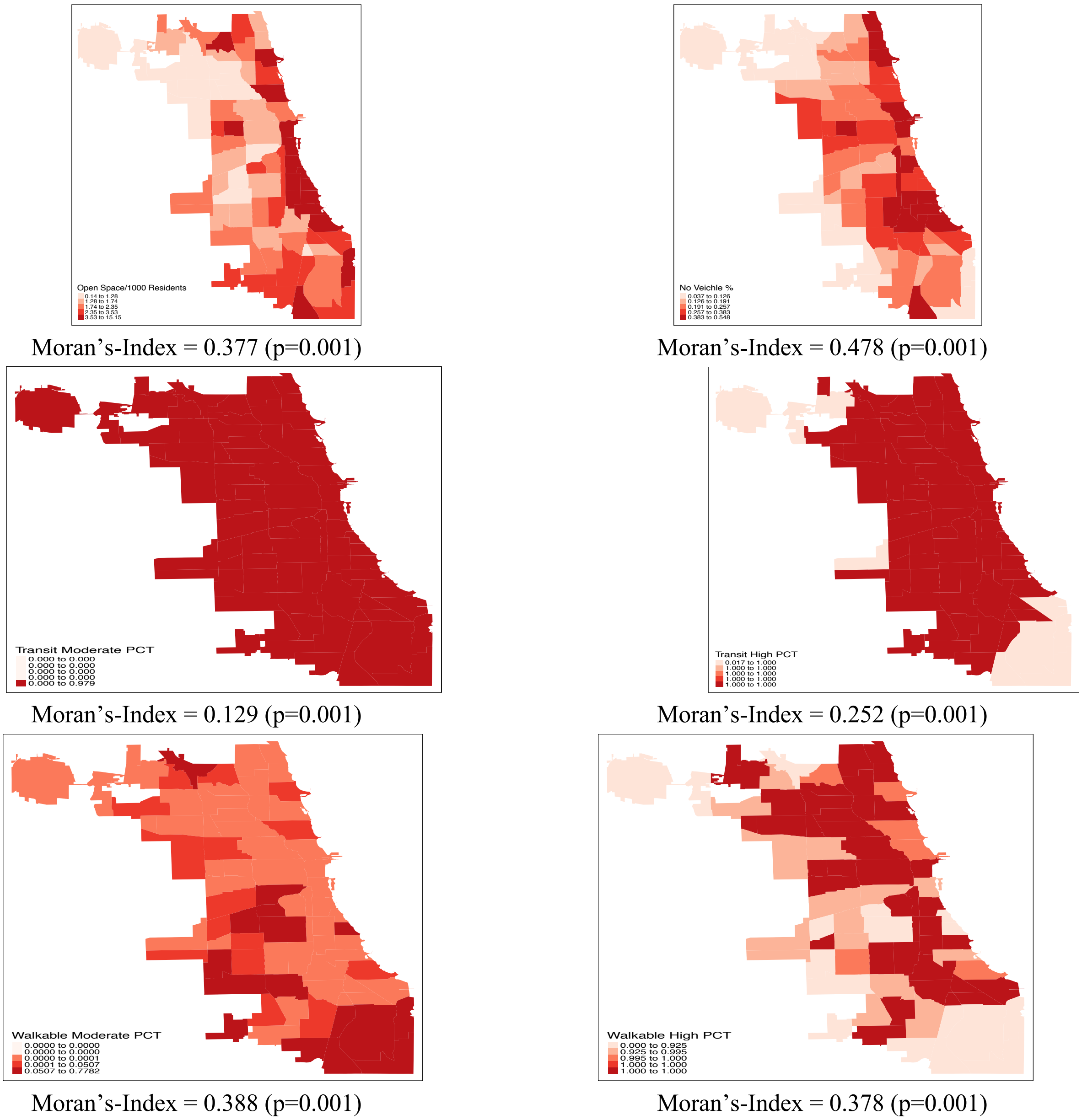

FIGURE 3. ***Spatial plot of the response and explanatory variables.***

### *B. Spatial Regression of the Total Rideshares in 2022*

To determine the statistically significant determinants for the total rideshares over the 77 areas, we estimate the spatial regression model in Equation (1). Because of multicollinearity, we employ a stepwise model selection procedure to identify the model with the best subset of significant predictors. Then, we assess the model's performance using the Deviance Information Criterion (DIC) [49] and the Watanabe-Akaike Information Criterion (WAIC) [50]. Table III summarizes the estimates of the spatial regression model with only significant predictors. The estimated model has DIC and WAIC 59.18 and 52.45, respectively.

Before exploring the results in Table III, we examine the residuals of the spatial regression model to identify any remaining spatial dependencies. Figure (4) shows no discernible leftover spatial residuals for the spatial regression model. Additionally, Moran's I for the residuals is almost negligible at -0.126 (p=0.97), assessing the model efficiency and accuracy in handling spatial dependence.

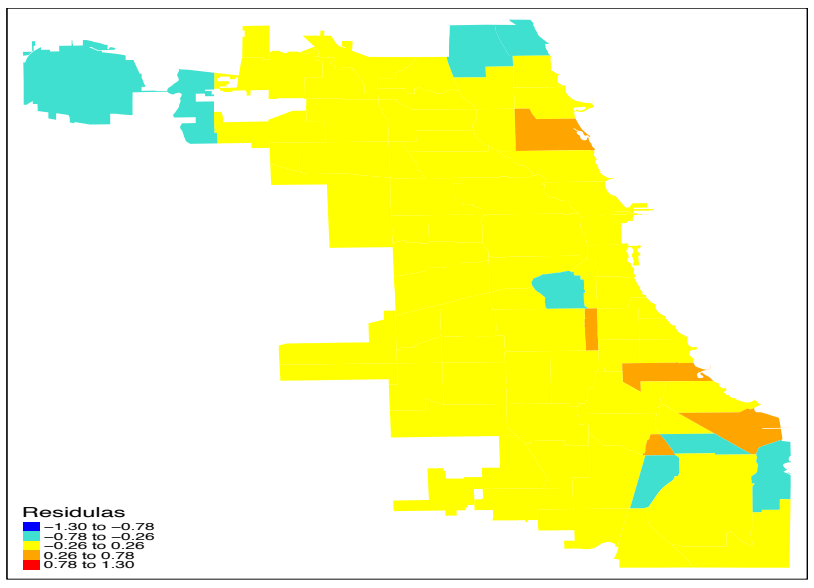

FIGURE 4. ***Spatial plot of residuals of the spatial model.***

The fitted model in Table III estimates a relatively high spatial dependence ($\rho = 0.611$). The estimated unstructured error variance is $\nu^2 = 0.106$, smaller than the spatial variance ($\tau^2 = 0.148$). TABLE III shows the estimated coefficients and 95% credible intervals, highlighting the importance of the demographic, economic, and transit factors in determining rideshares in Chicago.

TABLE III
POSTERIOR ESTIMATES OF THE BEST SPATIAL REGRESSION MODEL FOR CHICAGO RIDESHARE DETERMINANTS.

| | Posterior Estimates | | |
|---|---|---|---|
| Predictor | Mean | 2.5% Quantile | 97.5% Quantile |
| Intercept | -7.567 | -12.120 | -2.990 |
| Population 2020 | 0.423 | 0.056 | 0.787 |
| Crimes 2022 | 0.630 | 0.298 | 0.968 |
| Economically Active Percent | 0.351 | 0.035 | 0.664 |
| Median Income | 1.028 | 0.589 | 1.460 |
| No Vehicle Percent | 4.280 | 2.987 | 5.560 |
| Transit High PCT | 2.191 | 0.751 | 3.604 |
| Walkable Moderate PCT | -5.106 | -6.662 | -3.554 |
| Walkable High PCT | -4.195 | -5.456 | -2.953 |
| $\tau^2$ | 0.148 | 0.081 | 0.255 |
| $\nu^2$ | 0.106 | 0.069 | 0.158 |
| $\rho$ | 0.611 | 0.203 | 0.947 |

Demographic factors, particularly population size, have emerged as statistically significant determinants. As population size increases, ridesharing demand also increases. A second demographic determinant, total registered crimes in 2022, positively influences rideshares despite its correlation with population size.

Among the socioeconomic factors, economically active percentage, median income, and percentage of residents with no cars are statistically significant and positively affect ridesharing demand. Thus, rideshare determinants extend beyond population size to encompass the socioeconomic structure. The three transit-related factors are the final determinants. Specifically, the percentage of the high transit index category positively affects rideshares, whereas the percentage of moderate and high categories of the walkability index negatively influences rideshares in Chicago.

### *C. Spatio-temporal Regression of the Daily Rideshares in 2022*

Here, we fit the responses as daily rideshares over 77 Chicago areas. We consider the significant predictors identified in the previous spatial regression model TABLE III. In addition, we consider two new explanatory variables. The first is a time trend, to capturing potential long-term patterns in rideshare demand, especially after the Covid-19 pandemic. The second introduced predictor is an indicator/dummy variable for weekends, serving to assess the impact of weekends on rideshare demand.

The estimated spatiotemporal regression model in Table IV has overwhelming fitting performance in terms of DIC and WAIC. where DIC is -4659 and WAIC is -4537 despite the large penalties applied to them (3612 and 3341, respectively). According to the posterior mean estimates, there is strong spatial dependence as $\rho_S \approx 1$. In addition, the significance of $\rho_{1T}$ and $\rho_{2T}$ validates the choice of the second-order autoregressive (AR2) for the temporal residuals. The AR2 components account for potential daily seasonality in the residuals. Furthermore, there is reduction in the estimated spatial variance $\tau^2$ (0.016) and model error variance $\nu^2$ (0.044) compared to the corresponding variance components in the spatial regression model TABLE III. This illustrates the importance of implementing spatio-temporal dependence in model fitting.

TABLE IV
POSTERIOR ESTIMATES OF THE SPATIOTEMPORAL REGRESSION MODEL FOR CHICAGO RIDESHARE DETERMINANTS

| | Posterior Estimates | | |
|---|---|---|---|
| Predictor | Mean | 2.5% Quantile | 97.5% Quantile |
| Intercept | -12.988 | -13.539 | -12.440 |
| Daily Trend | 0.030 | 0.025 | 0.035 |
| Weekend Days | 0.029 | 0.021 | 0.039 |
| Population 2020 | 0.986 | 0.969 | 1.003 |
| Daily Crimes | 0.066 | 0.060 | 0.072 |
| Economically Active | 0.110 | 0.070 | 0.147 |
| Median Income | 0.966 | 0.907 | 1.022 |
| No Vehicle Proportion | 5.024 | 4.878 | 5.179 |
| Transit High PCT | 2.484 | 2.305 | 2.649 |
| Walkable Moderate PCT | -7.083 | -7.277 | -6.885 |
| Walkable High PCT | -4.937 | -5.059 | -4.810 |
| $\tau^2$ | 0.016 | 0.014 | 0.018 |
| $\nu^2$ | 0.044 | 0.043 | 0.045 |
| $\rho_S$ | 0.998 | 0.998 | 0.999 |
| $\rho_{1T}$ | -0.172 | -0.232 | -0.114 |
| $\rho_{2T}$ | *0.292* | *0.239* | *0.343* |

Our results reveal a positive trend (0.030) in daily rideshares in Chicago throughout 2022. This long-term trend exhibits a recovery from the decline in rideshares owing to Covid-19 closures and restrictions in 2020 and 2021. The estimated trends vary spatially through the 77 Chicago regions. This demonstrates the flexibility of employing our Bayesian hierarchical spatio-temporal framework and its flexibility compared to the classical SEM. Additionally, ride-sharing demand experienced a significant increase on weekends in Chicago, as evidenced by the estimated coefficient of 0.029. Finally, even after adjusting for the daily trend effect, the registered crimes are still statistically significant. This indicates that crime rate positively influences ride-sharing demand in Chicago, without the need for a proxy, such as the time trend.

Figure 5 shows a strong positive correlation between the actual rideshares and the fits of the spatiotemporal model (r = 0.98). Moreover, the slope in Figure 5 at 1.004 indicates a close alignment between the fit and the actual values.

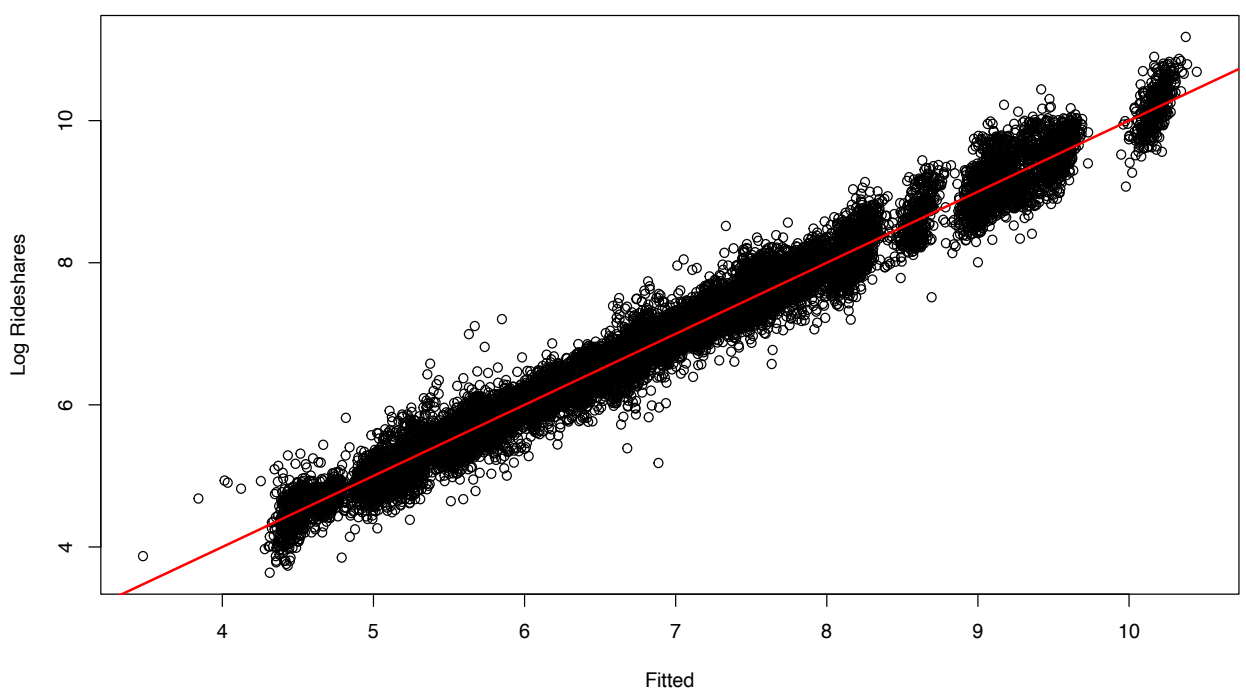

FIGURE 5. ***Scatterplot of the actual daily rideshares vs. fitted from the spatiotemporal model.***

Identification of Significant Variables: Our analysis identified several key variables that significantly influence the demand for ride-sharing in Chicago. These include population density, median income, and the number of transit stops within each community area.

Population Density: Higher population density was found to correlate strongly with increased ride-sharing demand. This is likely due to the greater need for transportation options in densely populated areas and the reduced feasibility of private car ownership.

Median Income: This variable showed a complex relationship with ride-sharing demand. Higher median income levels were associated with increased demand in some areas but decreased in others, suggesting that economic factors may influence the preference for different types of shared mobility services.

Transit Stops: The presence of more transit stops within an area was associated with higher ride-sharing demand, indicating that ride-sharing serves as a complement to public transit, particularly in facilitating the first and last mile of commuter journeys.

## V. Discussion

In this study, ride-sharing demand in Chicago was modeled using two different spatial and spatio-temporal models. Seven different variables are studied, and the results are discussed in this section.

### *A. Ride-sharing and population*

Figures 2 and 3 show that the spatial plot of ride-sharing demand and population density has a significant correlation. One common aspect of the impact of the population variable on ride-sharing is the variation in spatial distribution.

The North Side has become a vibrant center of activity and a sophisticated urban area. Known for its affluent residential areas, prosperous commercial zones, and artistic landmarks, it holds a crucial position in a city's dynamics. The region hosts notable attractions such as the Lincoln Park Zoo, Wrigley Field, and the Chicago History Museum, attracting both local inhabitants and a substantial influx of tourists, thereby increasing the need for taxi and ride-hailing services. An examination of the socio-demographic and land-use statistics of the North Side provides intriguing revelations. The population of this area in Chicago is significant, and it is known to have a higher income level and a more prosperous way of life than other areas in the city. The allocation of land for residential use, namely for single-family dwellings, is significantly greater in this area, indicating a lower population density in comparison to the South Side.

The ownership of vehicles in the North Side presents a contrasting depiction compared to the South Side. The area's higher economic level and apparent desire for personal transportation were reflected in the lower percentage of families without a vehicle. Nevertheless, despite this prevailing pattern, there continues to be a substantial need for ride-hailing services, possibly influenced by the convenience element and the vibrant nightlife and entertainment scenes in the area.

### *B. Ride-sharing and crime*

The Chicago Data Portal (CDP) provides crime records for several types of crime from 2000 to the current year. The crime variables represent the cumulative number of data records for the period from January 2015 to April 2019, as there are likely both lag and immediate effects of crime on ride-sharing usage. Hence, the crime variables are time invariant in the model. Figures 2 and 3 show a high correlation between population and crime variables. The higher the population, the higher the crime rate and associated with associated with a ride-sharing demand. This result

### *C. Ride-sharing and walkability*

The spatial model incorporated a walkability index. The walkability index represents the distance walked to parking lots or transit spaces[51], [52] . Our spatial model incorporated moderate and high walkability indices. The results indicated that a higher walkability index was associated with either low transit coverage, no parking space, or a higher parking rate. The findings suggest that (i) regions with limited transit options (such as no rail stations and only one or two bus stops, if any) result in a minimal number of ride-sharing trips; (ii) regions with a moderate level of transit service (such as a single rail station and numerous bus stops) generate the highest number of ride-sharing trips; and (iii) a small number of census tracts with extensive rail transit services (such as three or four rail stations) tend to generate significantly fewer ride-sharing trips compared to a similar census tract with no more than one rail station. Our findings suggest a positive correlation between the number of parking spaces and/or parking prices in each area and the walkability index. This indicates that areas with more parking spaces and higher parking rates tend to have a higher walkability index because people in these areas are more likely to utilize their private vehicles for transportation.

From a planning standpoint, all other things being equal, changing the availability of parking spaces and the fees charged for parking in the entire region can have a significant effect on the demand for ridesharing services to and from particular locations. From the standpoint of curbing management, the demand for ride-sharing services and the number of pickups and drop-offs are expected to increase as the availability of parking spaces declines and/or the parking cost increases.
From a transit planning standpoint, this suggests that transit can surpass ride-sharing in certain urban regions when the quality and availability of rail transit are exceptionally high.

### *D. Weekend vs. Weekday*

The binary variable representing weekends exhibited a positive and statistically significant relationship in both models, suggesting that weekends are associated with a higher number of ride-sharing trips. This phenomenon is expected, as a greater number of individuals engage in recreational and leisure pursuits on weekends. Additionally, past research indicates that the majority of consumers utilize ride-sharing services for social and leisure purposes [9], [35].

### *E. Economical characteristics*

High-income drivers generally have more trips per day because a ride-sharing business model often offers multiple bonuses when trip frequency targets are reached[53] . Researchers have indicated that young, well-educated, and higher-income individuals are higher users of ride-sharing [54]. Further noted that ride-sharing service users tend to be middle-income, young commuters, further revealing the income-related patterns in ride-sharing utilization. the expectation is that higher income census tracts will have higher ride-sharing usage this result is matching with [37], [40].

### *F. Car ownership and commute characteristics*

Our model indicates that ride-sharing demand is affected by car ownership. Tracts with less vehicle growth in ridesharing services may reduce the need for automobile ownership and dependence on cars [55], [56]. They also concluded that some households that owned more cars had a higher demand for ridesourcing services [12]. In contrast, people living in densely populated areas are more likely to use ride-sharing. This suggests that the link between ride-sharing and car ownership is contingent on the nature of the urban clusters. In addition, the potential for ride-sharing to lower private car ownership also depends on the extent to which it disrupts urban mobility services [57].

## VI. Policy implications and conclusions

This section first provides recommendations for ride-sharing companies to improve their daily operations and coordinate their shared responsibility for urban mobility based on our analysis results. We then summarize the key findings of this study.

### *A. Policy implications*

With the emergence of ride-sharing services, the public's travel options have expanded, and cutting-edge technologies have been successfully used. By utilizing these technological developments, traditional cab businesses can innovate in their sectors. The discussion that follows provides thorough recommendations for improving cooperative and effective services before, during, and after ride services based on our statistical model results.

Urban Planning and Population Density: The correlation between ride-sharing demand and population density, particularly in places such as the North Side of Chicago, shows the need for custom urban planning. In densely populated and prosperous areas, because of residential, commercial, and tourist activities, more rides are shared, and the transportation infrastructure needs to be enhanced. This might involve setting up more efficient traffic management systems and providing space for ride-sharing pickups and drops. In addition, creating new services, such as providing more charging stations for electrical vehicles helps to enhance transportation services. Such a system may require dynamically directed resource flows or extensive traffic control measures on weekends to deal with the rush.

Crime and Ridesharing: Higher crime rates are associated with increased ride-sharing demand, which could be a sign that certain places are dangerous, indicating a shift towards a broken-windows approach to transportation planning. Enhanced safety measures for ridesharing services in high-crime areas. In addition, rideshare data were incorporated into public safety strategies.
Walkability, Transit, and Parking: The relationship among walkability, transit availability, and ridesharing indicates that urban transit planning should weigh walkability against parking availability and public transit. The addition of rideshare can help areas with limited transit choices.

Conversely, regions with substantial transit coverage may have less demand for ride-sharing. Adjusting parking availability and rates could also affect ride-sharing demands, which require careful urban design and policy decisions. Adopting Transit-Oriented Demand (TOD) requires addressing various challenges, such as coordination among stakeholders, financing mechanisms, and equitable development practices. Overall, transit-oriented development represents a valuable and effective strategy for addressing the complex challenges of urbanization and creating more sustainable and inclusive cities.

Economic Characteristics and Ridesharing: Higher usage of ridesharing services by young, well-educated, and high-income people suggests that the use of such services could focus on these sectors. Policies might be tailored to encourage ride sharing in high-income areas while maintaining equitable access among all income levels.

Car ownership and ride-sharing: The complex relationship between car ownership and ride-sharing demands calls for a subtle approach to transportation policy. In areas with low car ownership, ride-sharing can be seen as a dominant form of transport. On the other hand, in areas with high car ownership, ride-sharing is best performed to assist personal vehicles, especially for those who want to turn away from car dependence.

Implications of Significant Variables: The influence of these significant variables provides insights into urban planning and transportation policy:
Urban Density and Transportation Planning: The positive correlation between population density and ride-sharing demand underscores the importance of integrating ride-sharing into the planning of transportation networks in urban areas. Urban planners might consider policies that support ride-sharing as part of a multimodal transportation strategy, especially in high-density areas.
Economic Factors and Service Targeting: The variable impacts of median income on ride-sharing demand suggest the need for differentiated marketing and service offerings across different income groups. For higher-income areas, premium ride-sharing services could be marketed more aggressively, whereas in lower-income areas, affordability and accessibility should be prioritized.
Integration with Public Transit: The relationship between transit stops and ride-sharing demand highlights the potential for coordinated services that link ride-sharing with public transit systems to enhance overall mobility, especially in underserved areas.
Future Research Directions: Further research could explore the causal relationships between these variables and ride-sharing demand by employing longitudinal data and experimental designs. Additionally, investigating these relationships in different urban settings could provide a more generalized understanding of how these factors influence shared mobility globally.

### *B. Conclusion and recommendation*

This study aimed to detect the determinants of ride-sharing demand in 77 community areas in Chicago. We considered three groups of explanatory variables: demographic, socioeconomic, land-use, and transit factors. The realizations of the factors considered under investigation showed a strong spatial correlation. Therefore, to determine the significant determinants of ride-sharing, we employ two Bayesian hierarchical models. These models account for both unstructured, random, errors, and structured, spatial, and spatiotemporal correlations.

The first model employs Bayesian spatial regression to analyze the determinants of total ride-share in 2022. The estimated model exhibits a small error variance, which is attributed to the spatial variance. This enhances the statistical power to detect the significant ride-sharing determinants. The fitted model reveals that in addition to population size, the economic structure of residents is a crucial factor. Furthermore, the number of crimes has emerged as a significant factor in ride-sharing. In essence, the relative safety offered by ride-sharing services compared with public transportation proves to be a key determinant of ride-sharing in Chicago.

Moreover, the spatial regression model emphasizes transit and walkability indices as comprehensive measures for assessing commuting availability. This clarifies why certain predictors, such as land use characteristics and parking availability, are not statistically significant.
Then, we employ a spatio-temporal regression model to analyze the daily ride-shares throughout 2022. The estimated model indicated a very strong spatial correlation and a much smaller unstructured variance compared with the previous spatial regression. This makes the spatiotemporal model more powerful for detecting ride-sharing determinants. Our analysis detected an increasing trend in rideshare across 77 Chicago areas in 2022. This signifies recovery from the rideshare drops observed in 2020 and 2021 due to Covid-19 restrictions and closures.

We generate operational and planning guidelines to expand these the current Ride-sharing services based on contributing component and described policy implications for these key findings Our full suite of bespoke and personalized services ensures that they meet each community's unique individual commuter needs during their travel. Implementation of a multi-modal transportation system optimizes transfers. Additional ancillary services added to transportation scene, e.g., to-go packaging and courier delivery, expand driver revenue streams and service offerings available to passengers. Future research can proceed in several directions, First, our model predicts and describes usage patterns using relatively static and time-variant variables (i.e., socio-demographic and land-use features), hence, having greater flexibility with predictor variable selection. Regression models can explore dynamic predictors not considered here, e.g., traffic congestion, weather, travel seasons, trip purpose, among other considerations to name just a few. Second, alternatives to hourly trip volume in each zone as the dependent variable can be pursued, e.g., other proxy variable to travel demand; consider, for instance, OD-based trip volume.

This research has conducted a comprehensive examination of the factors influencing ride-sharing demand in Chicago using sophisticated spatial and spatiotemporal Bayesian models. Our results highlight the substantial impact of population characteristics, economic conditions, and urban development patterns on ride-sharing behaviors. Notably, we have demonstrated that elevated crime rates and socioeconomic diversity across Chicago's communities significantly affect ride-sharing usage, indicating areas where targeted strategies could enhance service effectiveness and safety.

Major Contributions:
1. Sophisticated Analytical Methods: Through the utilization of spatial and spatiotemporal Bayesian analyses, this study advances the methodological approach to investigating urban mobility, providing a detailed understanding of the intricate dynamics that shape ride-sharing demand.
2. Urban Planning Insights: The research offers empirical data that can inform strategic planning, particularly in improving connectivity for underserved regions and integrating ride-sharing services with existing public transit systems to address first and last-mile challenges.
3. Policy Formulation: Our findings support the development of tailored policies that consider local demographic and economic characteristics, fostering equitable and efficient urban transportation solutions.

Future Research Directions:
1. Relationship Between Crime and Socio-Economic Variables: Further studies are necessary to investigate the complex interplay between crime rates and socio-economic factors. Such research could employ multi-level modeling techniques to better comprehend how these elements interact to influence ride-sharing demand.
2. Extended Time Frame Studies: To evaluate the long-term consequences of policy changes and economic shifts on ride-sharing patterns, future investigations should consider longitudinal data analysis.
3. Influence of Technological Progress: As ride-sharing technology advances, subsequent research should examine the effect of new features such as autonomous vehicles and enhanced mobile application functionalities on user adoption and satisfaction.
4. Cross-City Comparisons: Extending this research to other urban areas with differing layouts and transport policies could help validate the generalizability of our results and enhance the comparative analysis of ride-sharing dynamics.
5. Conducting causal analysis to quantify the impact of each of the statistically significant predictors on the ridesharing demand[58]. In addition, conduct trend analysis for future forecasting [59] and multiscale spatiotemporal analysis for the ridesharing demand [60] (c). Moreover, consider both the ridesharing demand and supply simultaneously in a multivariate framework for efficient pricing policies [61], [62].

Based on the determinants of ride-sharing demand identified in our study, we propose several targeted policy recommendations to enhance urban mobility strategies:

**1. Dynamic Pricing Policies**

- **Incentivize Off-Peak Ride-Sharing:** Implement dynamic pricing strategies that encourage ride-sharing during off-peak hours to alleviate traffic congestion during rush hours. This can be supported by our findings that ride-sharing demand spikes at certain times, suggesting potential for demand management.

**2. Zoning and Land Use Planning**

- **Integrate Ride-Sharing into Zoning Regulations:** Adjust zoning regulations to accommodate and even encourage ride-sharing friendly infrastructures, such as designated pick-up/drop-off zones in high-demand areas identified in our spatial analysis.

**3. Public Transit Integration**

- **Enhance First and Last Mile Connectivity:** Utilize our study's insights into transit deserts to deploy targeted ride-sharing solutions that complement public transit, improving overall accessibility and reducing reliance on private vehicle use.

**4. Accessibility Improvements**

- **Focus on Under-Served Areas:** Direct subsidies or support for ride-sharing in areas with lower socioeconomic status but high demand for mobility services, ensuring equitable access to transportation.

**5. Safety and Security Enhancements**

- **Implement Safety Enhancements in High Crime Areas:** In response to the correlation between crime rates and ride-sharing usage, introduce enhanced safety measures for ride-sharing operations in areas with higher crime rates to improve passenger security.

**6. Environmental Impact**

- **Promote Eco-Friendly Ride-Sharing Options:** Encourage the use of electric or hybrid vehicles for ride-sharing fleets to reduce environmental impact, aligning with urban sustainability goals.

**7. Data Sharing and Transparency**

- **Foster Partnerships for Data Sharing:** Develop partnerships between ride-sharing companies and city governments to facilitate data sharing, enhancing the ability to monitor, plan, and respond to urban mobility needs dynamically.

These recommendations are crafted to leverage the comprehensive analysis provided by our study, aiming to support urban planners and policymakers in making informed decisions that enhance the efficacy and sustainability of ride-sharing services within urban transportation ecosystems.

## REFERENCES

[1] S. Banerjee, B. P. Carlin, and A. E. Gelfand, Hierarchical modeling and analysis for spatial data. Chapman and Hall/CRC, 2003.

[2] J. LeSage and R. K. Pace, Introduction to spatial econometrics. Chapman and Hall/CRC, 2009.

[3] S. Sahu, Bayesian modeling of spatio-temporal data with R. CRC Press, 2022.

[4] D.-M. Storch, M. Timme, and M. Schröder, "Incentive-driven transition to high ride-sharing adoption," Nat Commun, vol. 12, no. 1, p. 3003, 2021.

[5] B. Kelleny and S. Ishak, "Spatiotemporal Analysis of the Relationships Between Pricing Strategies and Willingness to Share for Ridesourcing Modes," Transp Res Rec, vol. 2678, no. 1, pp. 393–409, 2024.

[6] R. Tachet et al., "Scaling law of urban ride sharing," Sci Rep, vol. 7, no. 1, pp. 1–6, 2017.

[7] Z. Li, "Understanding Factors Influencing Willingness to Ridesharing Using Big Trip Data and Interpretable Machine Learning," 2022, doi: 10.31219/osf.io/chy4p.

[8] R. Gangadharaiah et al., "A user-centered design exploration of factors that influence the rideshare experience," Safety, vol. 9, no. 2, p. 36, 2023.

[9] R. Mitra and J. T. Schofield, "Biking the First Mile: Exploring a Cyclist Typology and Potential for Cycling to Transit Stations by Suburban Commuters," Transportation Research Record Journal of the Transportation Research Board, vol. 2673, no. 4, pp. 951–962, 2019, doi: 10.1177/0361198119837229.

[10] M. Mahmoudi and X. Zhou, "Finding optimal solutions for vehicle routing problem with pickup and delivery services with time windows: A dynamic programming approach based on state–space–time network representations," Transportation Research Part B: Methodological, vol. 89, pp. 19–42, 2016.

[11] A. Tafreshian, N. Masoud, and Y. Yin, "Frontiers in service science: Ride matching for peer-to-peer ride sharing: A review and future directions," Serv Sci, vol. 12, no. 2–3, pp. 44–60, 2020.

[12] X. Zhang, Y. Xiang, Z. Zhou, Y. Xu, and X. Zhao, "Examining Spatial Heterogeneity of Ridesourcing Demand Determinants With Explainable Machine Learning," 2022, doi: 10.48550/arxiv.2209.07980.

[13] A. König and J. Grippenkoven, "Travellers' willingness to share rides in autonomous mobility on demand systems depending on travel distance and detour," Travel Behav Soc, vol. 21, pp. 188–202, 2020.

[14] X. Zhang, C. Shao, B. Wang, S. Huang, X. Mi, and Y. Zhuang, "Exploring the Role of Shared Mobility in Alleviating Private Car Dependence and On-Road Carbon Emissions in the Context of COVID-19," Front Environ Sci, vol. 10, p. 931763, 2022.

[15] P. Loa and K. N. Habib, "Identifying the determinants of anticipated post-pandemic mode choices in the Greater Toronto area: a stated preference study," Transp Res Rec, vol. 2677, no. 6, pp. 199–217, 2023.

[16] Y. Ma, B. Yu, and M. Xue, "Spatial Heterogeneous Characteristics of Ridesharing in Beijing–Tianjin–Hebei Region of China," Energies (Basel), 2018, doi: 10.3390/en11113214.

[17] F. Bistaffa, C. Blum, J. Cerquides, A. Farinelli, and J. A. Rodríguez-Aguilar, "A Computational Approach to Quantify the Benefits of Ridesharing for Policy Makers and Travellers," Ieee Transactions on Intelligent Transportation Systems, 2021, doi: 10.1109/tits.2019.2954982.

[18] C. N. Morrison, C. A. Mehranbod, M. Kwizera, A. Rundle, K. M. Keyes, and D. K. Humphreys, "Ridesharing and Motor Vehicle Crashes: A Spatial Ecological Case-Crossover Study of Trip-Level Data," Injury Prevention, vol. 27, no. 2, pp. 118–123, 2020, doi: 10.1136/injuryprev-2020-043644.

[19] S. Ali, G. Wang, and S. Riaz, "Aspect Based Sentiment Analysis of Ridesharing Platform Reviews for Kansei Engineering," Ieee Access, vol. 8, pp. 173186–173196, 2020, doi: 10.1109/access.2020.3025823.

[20] S. Gupta and J. V George, "App-Based Ride-Sharing Adoption Behaviour of Commuters: Evaluating Through TAM Approach," Orissa Journal of Commerce, 2022, doi: 10.54063/ojc.2022.v43i04.09.

[21] M. Zhang, M. Yang, D. Lei, and X. Song, "Bus Scheduling of Overlapping Routes Based on the Combination of All-Stop and Stop-Skipping Services," Transportation Research Record Journal of the Transportation Research Board, 2022, doi: 10.1177/03611981221098392.

[22] P. S. Lavieri, F. F. Dias, N. R. Juri, J. Kuhr, and C. R. Bhat, "A Model of Ridesourcing Demand Generation and Distribution," Transportation Research Record Journal of the Transportation Research Board, 2018, doi: 10.1177/0361198118756628.

[23] J. M. Sarriera, G. E. Álvarez, K. Blynn, A. Alesbury, T. R. Scully, and J. Zhao, "To Share or Not to Share," Transportation Research Record Journal of the Transportation Research Board, 2017, doi: 10.3141/2605-11.

[24] S. Rafi and A. N. Nithila, "Who Rides Uber Anyway? A Census-Tract Level Analysis and Clustering of Ride-Shares for the City of Chicago During the Era of the Pandemic," 2022, doi: 10.36227/techrxiv.21076042.

[25] T. Hansen and I. N. Sener, "Strangers on This Road We Are On: A Literature Review of Pooling in on-Demand Mobility Services," Transportation Research Record Journal of the Transportation Research Board, 2022, doi: 10.1177/03611981221123801.

[26] A. Brown, "Not All Fees Are Created Equal: Equity Implications of Ride-Hail Fee Structures," 2021, doi: 10.31219/osf.io/cpsqu.

[27] Y. Hou, V. Garikapati, D. Weigl, A. Henao, M. Moniot, and J. Sperling, "Factors Influencing Willingness to Pool in Ride-Hailing Trips," Transportation Research Record Journal of the Transportation Research Board, 2020, doi: 10.1177/0361198120915886.

[28] D. Xu, "The Complementary Effect of Ride-Sharing on Public Transit: Evidence From a Natural Experiment," Industrial Management & Data Systems, 2023, doi: 10.1108/imds-08-2022-0487.

[29] R. A. Mucci and G. D. Erhardt, “Understanding Ride-Hailing Sharing and Matching in Chicago Using Travel Time, Cost, and Choice Models,” Transportation Research Record Journal of the Transportation Research Board, 2023, doi: 10.1177/03611981231173636.
[30] H. Abkarian, Y. Chen, and H. S. Mahmassani, “Understanding Ridesplitting Behavior With Interpretable Machine Learning Models Using Chicago Transportation Network Company Data,” Transportation Research Record Journal of the Transportation Research Board, vol. 2676, no. 2, pp. 83–99, 2021, doi: 10.1177/03611981211036363.
[31] J. Soria, Y. Chen, and A. Stathopoulos, “K-Prototypes Segmentation Analysis on Large-Scale Ridesourcing Trip Data,” Transportation Research Record Journal of the Transportation Research Board, 2020, doi: 10.1177/0361198120929338.
[32] X. Yu and S. Shen, “An Integrated Decomposition and Approximate Dynamic Programming Approach for On-Demand Ride Pooling,” IEEE Transactions on Intelligent Transportation Systems, vol. 21, no. 9, pp. 3811–3820, 2020, doi: 10.1109/TITS.2019.2934423.
[33] N. D. Chan and S. A. Shaheen, “Ridesharing in North America: Past, present, and future,” Transp Rev, vol. 32, no. 1, pp. 93–112, 2012.
[34] A. Simonetto, J. Monteil, and C. Gambella, “Real-Time City-Scale Ridesharing via Linear Assignment Problems,” Transp Res Part C Emerg Technol, 2019, doi: 10.1016/j.trc.2019.01.019.
[35] L. Rayle, D. Dai, N. Chan, R. Cervero, and S. Shaheen, “Just a better taxi? A survey-based comparison of taxis, transit, and ridesourcing services in San Francisco,” Transp Policy (Oxf), vol. 45, pp. 168–178, 2016.
[36] L. V Hung, N. N. Trung, P. P. T. Thu, and D. N. Thi, “Intention to Provide Ridesharing Services: Determinants From the Perspective of Driver-Partners in a Gig Economy,” Problems and Perspectives in Management, 2022, doi: 10.21511/ppm.20(4).2022.24.
[37] A. Ghaffar, S. Mitra, and M. Hyland, “Modeling determinants of ridesourcing usage: A census tract-level analysis of Chicago,” Transp Res Part C Emerg Technol, vol. 119, Oct. 2020, doi: 10.1016/j.trc.2020.102769.
[38] Z. Wang, Y. Zhang, B. Jia, and Z. Gao, “Comparative Analysis of Usage Patterns and Underlying Determinants for Ride-hailing and Traditional Taxi Services: A Chicago Case Study,” Transp Res Part A Policy Pract, vol. 179, Jan. 2024, doi: 10.1016/j.tra.2023.103912.
[39] N. Barbour, Y. Zhang, and F. Mannering, “An exploratory analysis of the role of socio-demographic and health-related factors in ridesourcing behavior,” J Transp Health, vol. 16, p. 100832, 2020.
[40] P. F. Belgiawan, T. B. Joewono, and M. Z. Irawan, “Determinant factors of ride-sourcing usage: A case study of ride-sourcing in Bandung, Indonesia,” Case Stud Transp Policy, vol. 10, no. 2, pp. 831–840, 2022.
[41] Á. Aguilera-García, J. Gomez, G. Velázquez, and J. M. Vassallo, “Ridesourcing vs. traditional taxi services: Understanding users’ choices and preferences in Spain,” Transp Res Part A Policy Pract, vol. 155, pp. 161–178, 2022.
[42] F. Feng, X. Li, Y. Guo, and C. Cheng, “Understanding factors that impact ridesourcing service usage frequency: a case study in Shanghai,” Transportation Planning and Technology, vol. 46, no. 4, pp. 462–481, 2023.
[43] M. Rizki, T. B. Joewono, P. F. Belgiawan, and M. Z. Irawan, “The travel behaviour of ride-sourcing users, and their perception of the usefulness of ride-sourcing based on the users’ previous modes of transport: A case study in Bandung City, Indonesia,” IATSS research, vol. 45, no. 2, pp. 267–276, 2021.
[44] Y. Xu, X. Yan, X. Liu, and X. Zhao, “Identifying key factors associated with ridesplitting adoption rate and modeling their nonlinear relationships,” Transp Res Part A Policy Pract, vol. 144, pp. 170–188, 2021.
[45] S. Sabouri, K. Park, A. Smith, G. Tian, and R. Ewing, “Exploring the influence of built environment on Uber demand,” Transp Res D Transp Environ, vol. 81, p. 102296, 2020.
[46] J. Besag, J. York, and A. Mollié, “Bayesian image restoration, with two applications in spatial statistics,” Ann Inst Stat Math, vol. 43, pp. 1–20, 1991.
[47] A. Rushworth, D. Lee, and R. Mitchell, “A spatio-temporal model for estimating the long-term effects of air pollution on respiratory hospital admissions in Greater London,” Spat Spatiotemporal Epidemiol, vol. 10, pp. 29–38, 2014.
[48] D. Lee, A. Rushworth, and G. Napier, “Spatio-temporal areal unit modeling in R with conditional autoregressive priors using the CARBayesST package,” J Stat Softw, vol. 84, pp. 1–39, 2018.
[49] D. J. Spiegelhalter, N. G. Best, B. P. Carlin, and A. Van Der Linde, “Bayesian measures of model complexity and fit,” J R Stat Soc Series B Stat Methodol, vol. 64, no. 4, pp. 583–639, 2002.
[50] S. Watanabe and M. Opper, “Asymptotic equivalence of Bayes cross validation and widely applicable information criterion in singular learning theory.,” Journal of machine learning research, vol. 11, no. 12, 2010.
[51] I. Jeong, M. Choi, J. Kwak, D. Ku, and S. Lee, “A comprehensive walkability evaluation system for promoting environmental benefits,” Sci Rep, vol. 13, no. 1, Dec. 2023, doi: 10.1038/s41598-023-43261-0.
[52] K. M. Leyden, M. J. Hogan, L. D’Arcy, B. Bunting, and S. Bierema, “Walkable Neighborhoods: Linkages Between Place, Health, and Happiness in Younger and Older Adults,” Journal of the American Planning Association, 2023, doi: 10.1080/01944363.2022.2123382.
[53] T. B. Joewono, M. Rizki, and J. Syahputri, “Does Job Satisfaction Influence the Productivity of Ride-Sourcing Drivers? A Hierarchical Structural Equation Modelling Approach for the Case of Bandung City Ride-Sourcing Drivers,” Sustainability, 2021, doi: 10.3390/su131910834.
[54] R. Li, “Joint Modeling of User Behaviors Based on Variable-Order Additive Markov Chain for POI Recommendation,” Wirel Commun Mob Comput, vol. 2021, pp. 1–13, 2021, doi: 10.1155/2021/4359369.
[55] A. R. Khavarian-Garmsir, A. Sharifi, and M. H. H. Abadi, “The Social, Economic, and Environmental Impacts of Ridesourcing Services: A Literature Review,” Future Transportation, 2021, doi: 10.3390/futuretransp1020016.
[56] H. Mostofi, “The Association Between ICT-Based Mobility Services and Sustainable Mobility Behaviors of New Yorkers,” Energies (Basel), 2021, doi: 10.3390/en14113064.
[57] M. J. Mohamed, T. Rye, and A. Fonzone, “Operational and Policy Implications of Ridesourcing Services: A Case of Uber in London, UK,” Case Stud Transp Policy, 2019, doi: 10.1016/j.cstp.2019.07.013.
[58] F. Li, P. Ding, and F. Mealli, “Bayesian causal inference: a critical review,” Philosophical Transactions of the Royal Society A, vol. 381, no. 2247, p. 20220153, 2023.
[59] M. Elkhouly, T. Alhadidi, and H. Rakha, “Modeling Traffic Vehicular Counts at an Urban Intersection Using Complex Seasonal Modeling Techniques,” KSCE Journal of Civil Engineering, vol. 29, no. 1, p. 100040, Jan. 2025, doi: 10.1016/j.kscej.2024.100040.
[60] M. A. R. Ferreira, S. H. Holan, and A. I. Bertolde, “Dynamic multiscale spatiotemporal models for Gaussian areal data,” J R Stat Soc Series B Stat Methodol, vol. 73, no. 5, pp. 663–688, 2011.
[61] M. Elkhouly and M. A. R. Ferreira, “Multiscale Heteroscedastic Multivariate Spatio-Temporal Models,” in Modeling Spatio-Temporal Data, Chapman and Hall/CRC, 2025, pp. 140–190.
[62] M. Elkhouly and M. A. R. Ferreira, “Dynamic multiscale spatiotemporal models for multivariate Gaussian data,” Spat Stat, vol. 41, p. 100475, 2021.

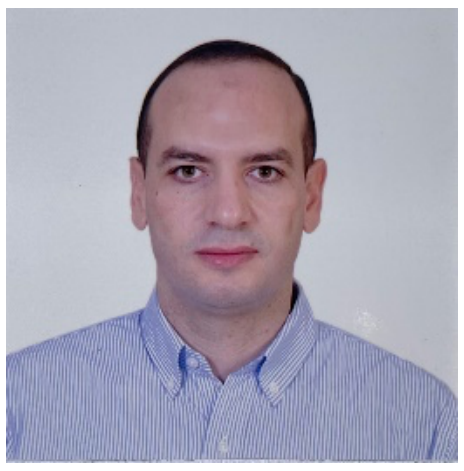

**MOHAMED ELKHOULY** (M'84) received a B.S. degree, majoring in statistics and minoring in economics, from the Faculty of Economics, Cairo University, Cairo, Egypt, in 2008. He earned his M.S. and Ph.D. degrees in statistics from Virginia Tech, Virginia, in 2014 and 2019, respectively.

From 2008 to 2012, he was a Teaching Assistant at Cairo University. He served as a Visiting Assistant Professor at the University of Wisconsin-Madison from 2019 to 2021 and worked as an Assistant Professor of statistics at Cairo University from 2021 to 2022. Since 2022, he has been an Assistant Professor in the Mathematics and Statistics Department at Cleveland State University. His research interests include spatio-temporal modeling, state-space models, Bayesian analysis, multivariate analysis, and generalized linear models.

Dr. Elkhouly received the Klaus Hinkelmann Award for outstanding service by a graduate student at Virginia Tech in 2017. Since 2014, he has been a member of Mu Sigma Rho, the National Statistics Honorary Society, and the American Statistical Association (ASA).

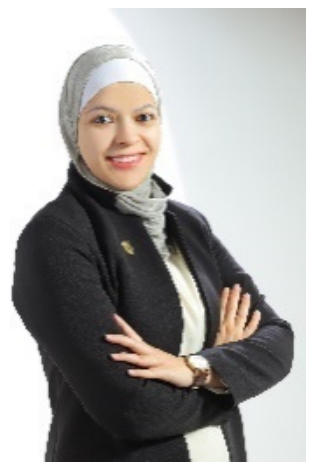

**TAQWA I.ALHADIDI** She received the B.S. degree in Highway and Bridges Engineering in 2012 and M.S. degrees in Transportation Engineering from the University of Jordan, in 2015 and the Ph.D. degree in civil engineering from Virginia Tech in Blacksburg, VA, USA in 2020.
From 2017 to 2020, she was a Research Assistant with the Center of Sustainable mobility in VTTI. From 2020 to June 2024, she has been an Assistant Professor with the civil Engineering Department, Al-Ahliyya Amman University, Amman, Jordan. Her research interests include transportation system optimization, modeling, and assessment. She also works on using data science algorithms in traffic safety and transit modeling. She is an active member in different professional committees including ASCE, TRB and WTS. She is also an active reviewer in different journals including ASCE, TRR, JJCE and TRIP.